\documentclass[journal]{IEEEtran}
\usepackage{cite}
\usepackage{amsmath,amssymb,amsfonts}
\usepackage{algorithmic}
\usepackage{graphicx}
\usepackage{textcomp}
\usepackage{xcolor}
\usepackage{float}
\usepackage{array}
\usepackage{booktabs}
\usepackage{subfigure}
\usepackage{amsmath}
\usepackage{multirow}

\usepackage{mathtools}
\usepackage{adjustbox}
\usepackage{tabulary}

\usepackage{algorithm}

\def\BibTeX{{\rm B\kern-.05em{\sc i\kern-.025em b}\kern-.08em
    T\kern-.1667em\lower.7ex\hbox{E}\kern-.125emX}}
\begin{document}

\title{Hardware-Robust In-RRAM-Computing for Object Detection\\
}
\bstctlcite{IEEEexample:BSTcontrol}
\author{\IEEEauthorblockN{Yu-Hsiang Chiang, Cheng En Ni, Yun Sung, Tuo-Hung Hou, \textit{Senior Member, IEEE}, Tian-Sheuan Chang, \textit{Senior Member, IEEE}, and  Shyh Jye Jou, \textit{Senior Member, IEEE}}
\thanks{Manuscript received Dec. 05, 2021; revised Feb. 14, and Mar. 26, 2022, and accepted Apr. 27, 2022.}
\thanks{This work was supported by TSMC and the Ministry of Science and Technology, Taiwan, under Grant 109-2634-F-009 -022,  109-2639-E-009-001 and 110-2622-8-009-018-SB, and 110-2218-E-A49-015-MBK. The authors are with the Institute of Electronics, National Yang Ming Chiao Tung University, Hsinchu 30010, Taiwan (e-mail: q123783293@gmail.com, nison.zxc@gmail.com, sung86084@gmail.com, thhou@mail.nctu.edu.tw, tschang@nycu.edu.tw, jerryjou@g2.nctu.edu.tw) }
\thanks{
© 2022 IEEE.  Personal use of this material is permitted.  Permission from IEEE must be obtained for all other uses, in any current or future media, including reprinting/republishing this material for advertising or promotional purposes, creating new collective works, for resale or redistribution to servers or lists, or reuse of any copyrighted component of this work in other works.\\
Y. -H. Chiang, C. E. Ni, Y. Sung, T. -H. Hou, T. -S. Chang and S. J. Jou, "Hardware-Robust In-RRAM-Computing for Object Detection," in IEEE Journal on Emerging and Selected Topics in Circuits and Systems, doi: 10.1109/JETCAS.2022.3171522.
}
}

\maketitle

\begin{abstract}
In-memory computing is becoming a popular architecture for deep-learning hardware accelerators recently due to its highly parallel computing, low power, and low area cost. However, in-RRAM computing (IRC) suffered from large device variation and numerous nonideal effects in hardware. Although previous approaches including these effects in model training successfully improved variation tolerance, they only considered part of the nonideal effects and relatively simple classification tasks. This paper proposes a joint hardware and software optimization strategy to design a hardware-robust IRC macro for object detection. We lower the cell current by using a low word-line voltage to enable a complete convolution calculation in one operation that minimizes the impact of nonlinear addition. We also implement ternary weight mapping and remove batch normalization for better tolerance against device variation, sense amplifier variation, and IR drop problem. An extra bias is included to overcome the limitation of the current sensing range. The proposed approach has been successfully applied to a complex object detection task with only 3.85\% mAP drop, whereas a naive design suffers catastrophic failure under these nonideal effects.

\end{abstract}


\maketitle

\section{Introduction}
In-memory computing (IMC) \cite{mfchang2020survey} has attracted significant attention in recent years to implement deep neural networks due to its highly parallel operation by using memory crossbar arrays and low power consumption based on analog computing. However, the inherent variation of analog computing makes the values of model weight and activation deviated from the original well-trained model and easily leads to degradation of model accuracy or even catastrophic failure. This situation becomes even worse for in-RRAM-computing (IRC) due to larger device variations of RRAM cells and other nonideal effects. 

To solve the above issues, one common approach is to model these nonideal effects during model training such that the trained model can more tolerate these nonideal effects. NIA\cite{fouda2020ir} showed the accuracy degradation caused by IR drop. The accuracy dropped to 32\% on a simple dataset such as MNIST even with a relatively small crossbar array size(64x64). This accuracy degradation could be recovered to 98.3\% by retraining networks with approximated IR drop noise. IR-QNN\cite{he2019noise} showed that the problems of IR drop, device variation, and driver nonlinearity could be solved by training with an average mask matrix that helps to predict the nonideal effects on BNN.
Lee et al. \cite{lee2020learning} also showed a method to recover the accuracy degradation due to IR drop by training the distorted vector-matrix-multiplication (VMM) when mapping binary weights to the crossbar array. Beyond these offline training methods, Chang et al. \cite{chang2019nv} combined offline variation-aware training and online fine-tuning or retraining part of the layers to reduce the impacts of device variation and recover accuracy.  

Most of the related works focus on modeling these nonideal effects in the training process. However, since these works only perform simple classification tasks and use a smaller crossbar array size ($<$256x256), the accuracy degradation could be easily recovered using the proposed training or retraining methods. By contrast, this paper targets a larger crossbar array size  (1024x1024), and a more complex task: object detection, which is much more sensitive to nonideal effects. Therefore, deploying only these offline training or retraining methods cannot recover the accuracy loss due to the nonideal effects completely.

This paper returns to a more fundamental strategy: making the IRC macro more robust hardware through joint hardware and software optimization. In hardware optimization, we reduce the RRAM cell current by using a lower word-line voltage to enable a complete convolution calculation in one operation rather than relying on external addition from multiple sub-convolution blocks, thus minimizing the nonlinear addition effect. In software (algorithm) optimization, we adopt ternary weight mapping without batch normalization (BN) to reduce the effects of device variation, sense amplifier (SA) variation, and IR drop. The limited range of current sensing is overcome using an extra bias. With these approaches, we do not require a lengthy training process as in other approaches while maintaining the mean Average Precision (mAP) loss of less than 4\% considering hardware nonideality even for the complex object detection task. This loss could be further recovered by a simple fine-tuning of the final layer.

The rest of the paper is organized as follows. Section II shows the overview of IMC and the measured distributions of RRAM resistance from an IRC test chip. Section III shows the nonideal effects of the IRC macro based on SPICE simulation. Based on the measured RRAM distribution and simulated IRC nonidealities, Section IV shows the proposed approach to tackle these nonideal effects for an object detection task. Section V presents the experimental results showing the effectiveness of our approaches to improve the model accuracy and comparison with other works. Finally, this paper is concluded in Section VI.

\section{Overview of IMC and RRAM Test Chip}

\subsection{Overview of in-RRAM Computing}

\begin{figure}[htb]
  \centering
  \includegraphics[height=!,width=1.0\linewidth,keepaspectratio=true]
  {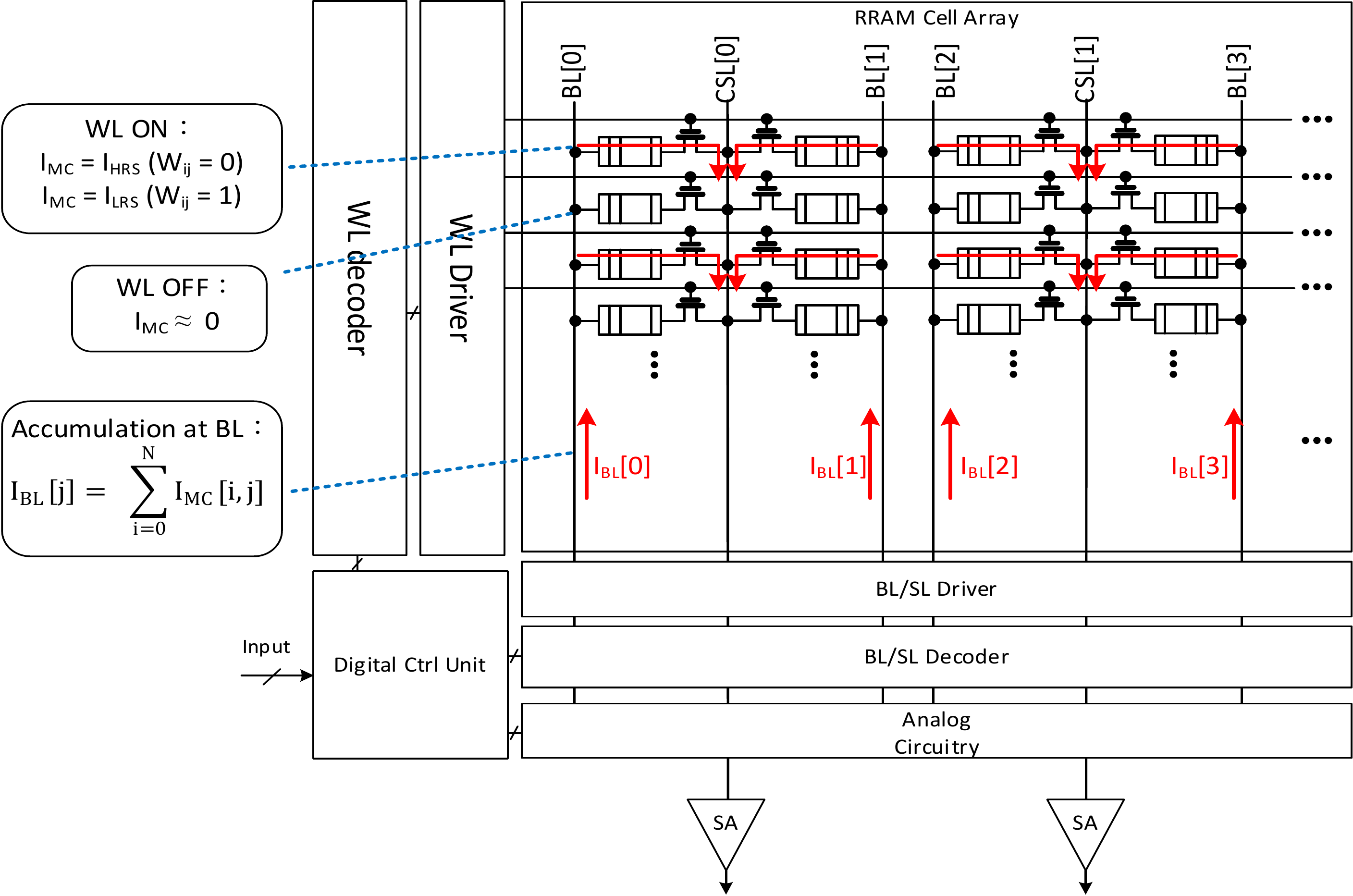}
  \caption[In-memory BN flow.]{{A typical IRC macro and its computing flow.}}
  \label{fig:rram_cim_flow}
\end{figure}

Fig.~\ref{fig:rram_cim_flow} shows a typical IRC macro and its computing flow for VMM used in convolution neural networks. VMM multiplies an input vector with the model weight matrix and accumulates the results as the convolution outputs. As shown in the figure, for IMC using RRAM, we store the weight ($w_i$) in the bit cell and apply the input feature ($x_i$) on the word-lines $WL$ to active/inactive cells. With the input of one, the cell current depends on the cell weight state. With the input of zero, the deactivated cell produces nearly zero current. When activating multiple $WL$, the current from bit cells of the same bit-line $BL$ is summed together as the VMM results. These results are then converted to a digital output using an analog-digital converter (ADC) for further processing. For a binary neural network (BNN), multi-bit ADCs are replaced by simple SAs to generate only binary activation.   

In this work, we focus on BNN and ternary weight neural network(TNN) \cite{courbariaux2016binarized, li2016ternary, alemdar2017ternary} as our target because of their compact models as well as the simple binary activation function. Binary activation provides the lowest possible hardware overhead on IRC peripheral circuits while maintaining good performance in various tasks. For BNN, both weights and activation are binary. For TNN, the weights are ternary ($0, \pm 1 $), and the activations are binary. To fit one convolution channel of neural networks to an IMC macro, the widely used BN layer \cite{ioffe2015batch} is integrated into the IMC macro by using an in-memory BN mapping method \cite{yonekawa2017chip}. The BN calculation is converted into a bias that is mapped into IRC as normal weights with fixed inputs. Thus, both BN and convolution are computed completely in the analog domain with no need of multi-bit intermediate results, followed by binary activation performed by SAs.

\subsection{RRAM Test Chip}
\begin{figure}[htb]
  \centering
  \includegraphics[height=!,width=0.7\linewidth,keepaspectratio=true]
  {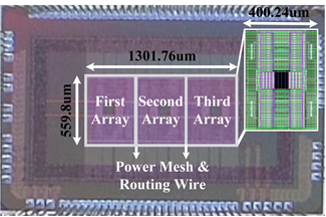}
  \centering
  \caption[.]{{The die photo of the RRAM test chip.}}
  \label{fig:rram_test_chip}
\end{figure}

\begin{figure}[htb]
  \centering
  \includegraphics[height=!,width=1.0\linewidth,keepaspectratio=true]
  {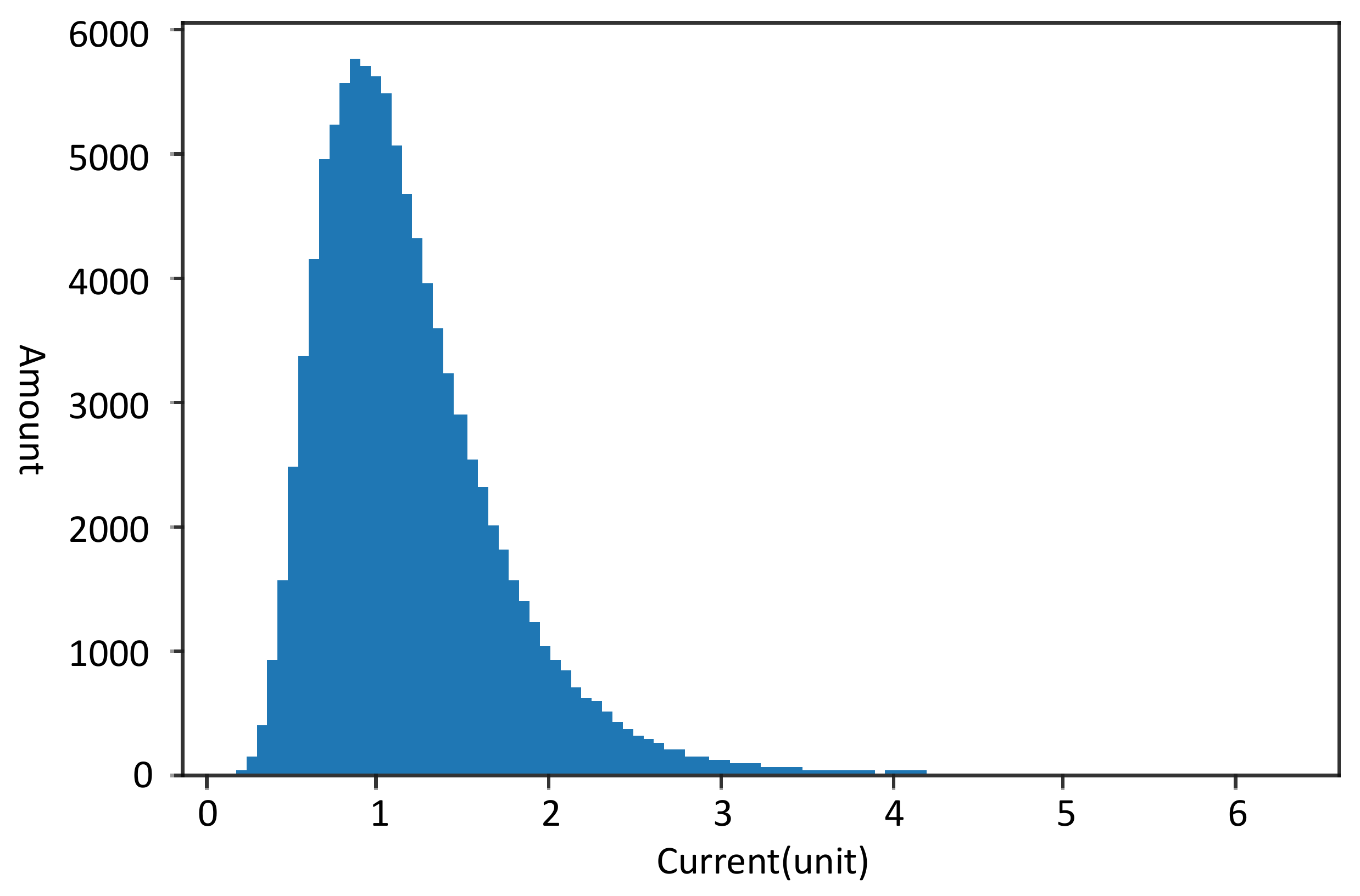}
  \caption[Current distribution of RRAM cells considering device variation.]{{Current distribution of RRAM cells considering device variation with 100k samples. The device is formed to LRS and biased at a word-line voltage of 0.44V.}}
  \label{fig:RRAM_device_variation_distribution}
\end{figure}

We have designed and implemented a RRAM test chip to characterize the device characteristics, as shown in Fig.~\ref{fig:rram_test_chip}. This chip implemented using TSMC 40nm CMOS plus embedded RRAM process \cite{TSMCRRAM, lee20171, chou2018n40} and included several 256x256 RRAM arrays for testing purposes. The RRAM device was designed for storage applications. Its typical characteristics include a forming voltage of around 3 V, set and reset voltages of around +1.5V and -1.5V, respectively. Typical resistance values of low-resistance state (LRS) and high-resistance state (HRS) from storage applications are around 10$^{4}$ ohms and 10$^{5}$ ohms, respectively \cite{TSMCRRAM, lee20171, chou2018n40}. However, because the nonstandard lower word-line voltage is used in our IRC macro design, the characteristics of RRAM cells including variations are obtained from this RRAM test chip. The typical resistance of the 1T1R RRAM cell at the word-line voltage of 0.44 V exceeds 10$^{5}$ ohms when 0.1 V is applied across the 1T1R RRAM cell, indicating that the resistance of the RRAM access transistor is higher than the typical LRS and HRS of RRAM for storage due to the low word-line voltage. Fig. ~\ref{fig:RRAM_device_variation_distribution} shows the typical resistance distribution. The standard deviation in this log-normal distribution is around 0.42. This state is used as the LRS for IRC. We chose the RRAM cell without forming with a resistance exceeding 10$^{9}$ ohms to represent the HRS for IRC. The variation of non-forming devices is negligible. These measured resistance and variation distributions are considered in our IRC macro design. The choice of lower word-line voltage and higher resistance states in IRC is critical for implementing a large-size IRC macro, which will be discussed next. Note that using the non-forming state to represent the HRS means that the model stored in IRC is one-time programmable, which could be used for inference applications requiring no model updates after deployment. For those applications requiring frequent model updates, the development of RRAM technology with higher resistance values would benefit the energy efficiency and accuracy of IRC in the future.


\section{Nonideal Effects of the IRC Macro}

\begin{figure}[htb]
  \centering
  \includegraphics[height=!,width=1.0\linewidth,keepaspectratio=true]
  {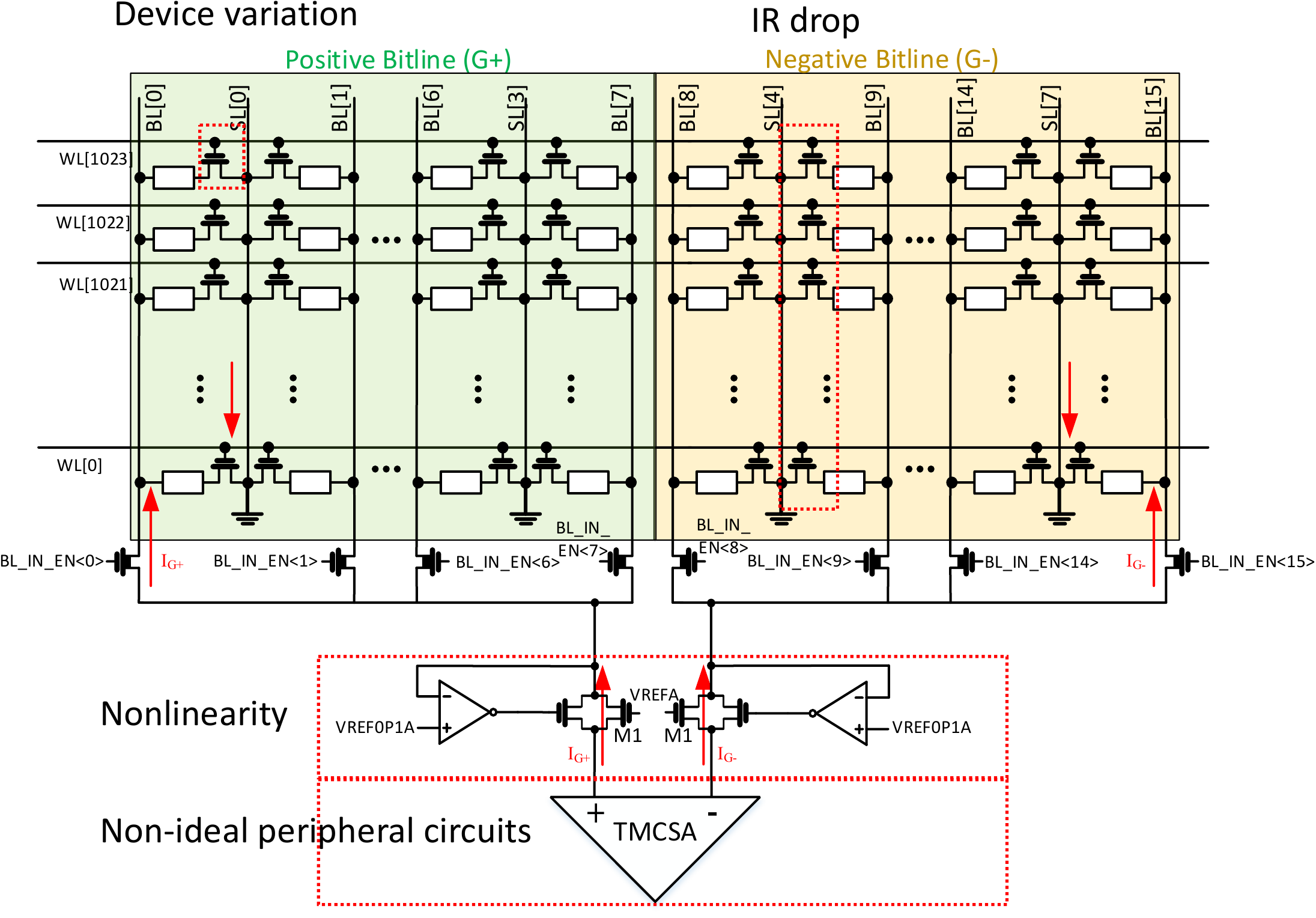}
  \caption{{Nonideal effects of the proposed IRC macro, including device variation, nonlinearity, IR drop, and nonideal peripheral circuits. Best viewed in colors. }}
  \label{fig:IRC_non_ideal}
\end{figure}

The nonideal effects of IRC can lead to a catastrophic failure for model inference because of the significant weight and activation deviations from the ideally-trained model. The causes of these nonideal effects include variations in RRAM cell conductance, nonlinear current accumulation, IR drop on the bit-lines, and nonideal peripheral circuits as shown in Fig.~\ref{fig:IRC_non_ideal}. The SA design is based on the triple-margin current-mode sense amplifier (TMCSA) from \cite{TMCSA} with additional switches to reduce the power consumption. Below we show these nonideal effects considering the measurement results from the test chip (Figs. 2 and 3) and a complete 1024x1024 IRC macro design simulated using SPICE and the TSMC 40nm embedded RRAM technology (Section III). The impact of these non-ideal effects on practical AI models and potential solutions are further discussed in Section IV and V. 

\subsection{Target Large-Sized IRC Macro}
\begin{figure}[htb]
  \centering
  \includegraphics[height=!,width=1.0\linewidth,keepaspectratio=true]
  {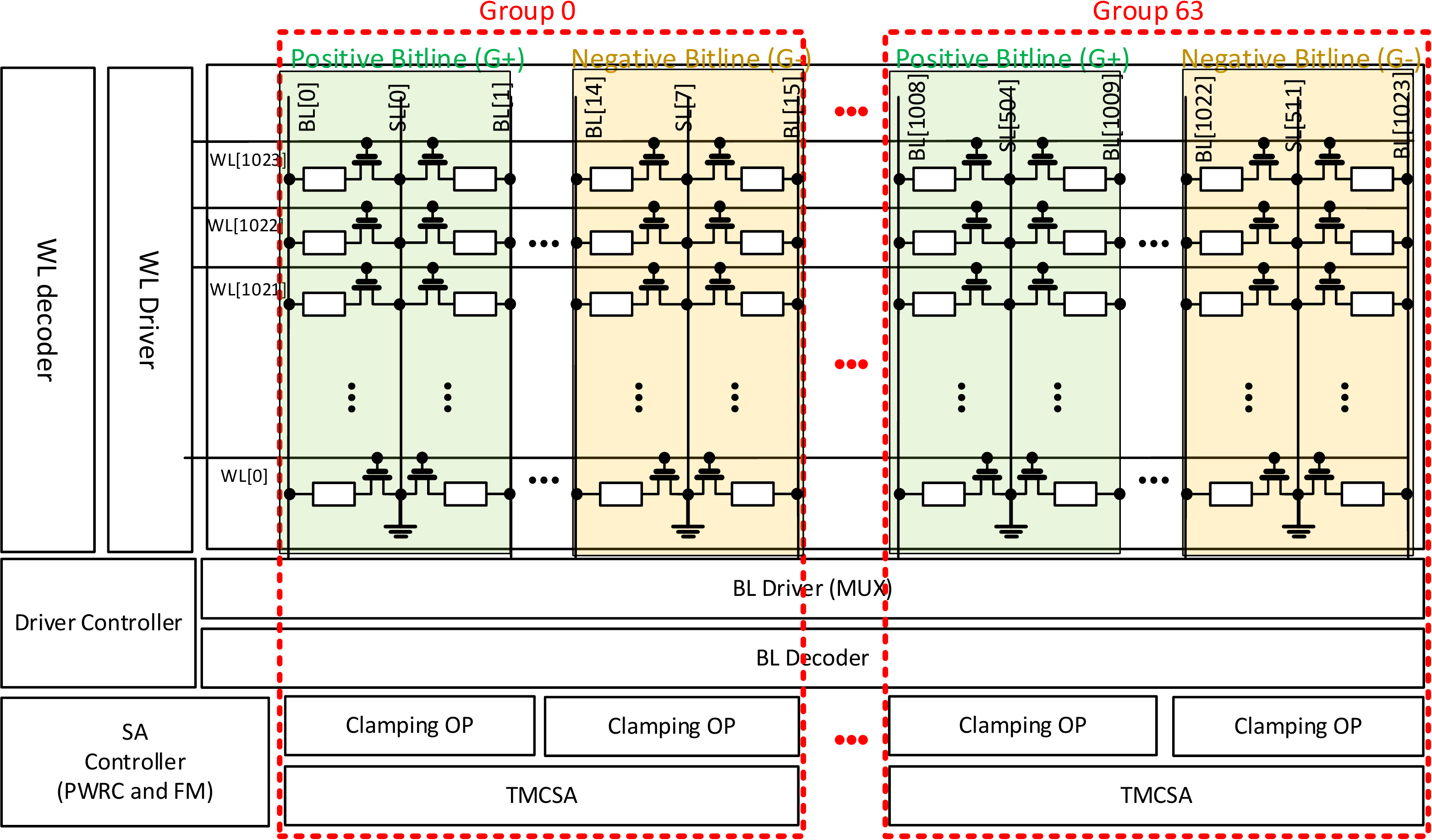}
  \caption{{The architecture of the proposed IRC macro. }}
  \label{fig:IRC_macro}
\end{figure}

This IRC macro includes one 1024x1024 RRAM array with IRC peripheral circuits as shown in Fig.~\ref{fig:IRC_macro}. In our IRC design, all 1024 word-lines are turned on simultaneously to multiply with the pre-stored weights in the RRAM cells, and the currents from each RRAM cell are accumulated on the bit-line as the analog convolution output. The convolution output is then converted to a binary activation through the binary SA. The binary SA is adopted to avoid using high-precision ADC conversion. Most other IMC designs try to emulate the conventional digital process by turning on only a small number of bit-lines to generate an accurate multibit output using ADC as a partial sum, followed by high-precision digital accumulation and activation calculation. The high-precision partial sum is necessary, otherwise, the accuracy of digital accumulation becomes substantial. This hybrid IMC-digital approach requires area- and power-expensive multibit ADCs and compromises parallelism for computation. By contrast, our design performs accurate analog accumulation on the same bit-line and directly outputs the final binary activation without partial-sum accumulation. As a result, we can fully exploit the high parallelism of a large-sized macro and avoid the quantization errors of partial sums at the same time. But this design has to consider the maximum current allowed in the RRAM array (300$\mu A$ in our case) and the IR drop problem that will be discussed later. To mitigate these constraints, lowering the RRAM cell current by using a lower word-line voltage is necessary for implementing a large-size IRC macro. Note that the reduced cell current would not prolong the SA sensing time, thus not affecting the operating frequency of the IRC macro. This is because the total bit-line current accumulated from 1024 RRAM cells remains high enough for fast SA sensing.

\begin{figure}[htb]
  \centering
  \includegraphics[height=!,width=1.0\linewidth,keepaspectratio=true]
  {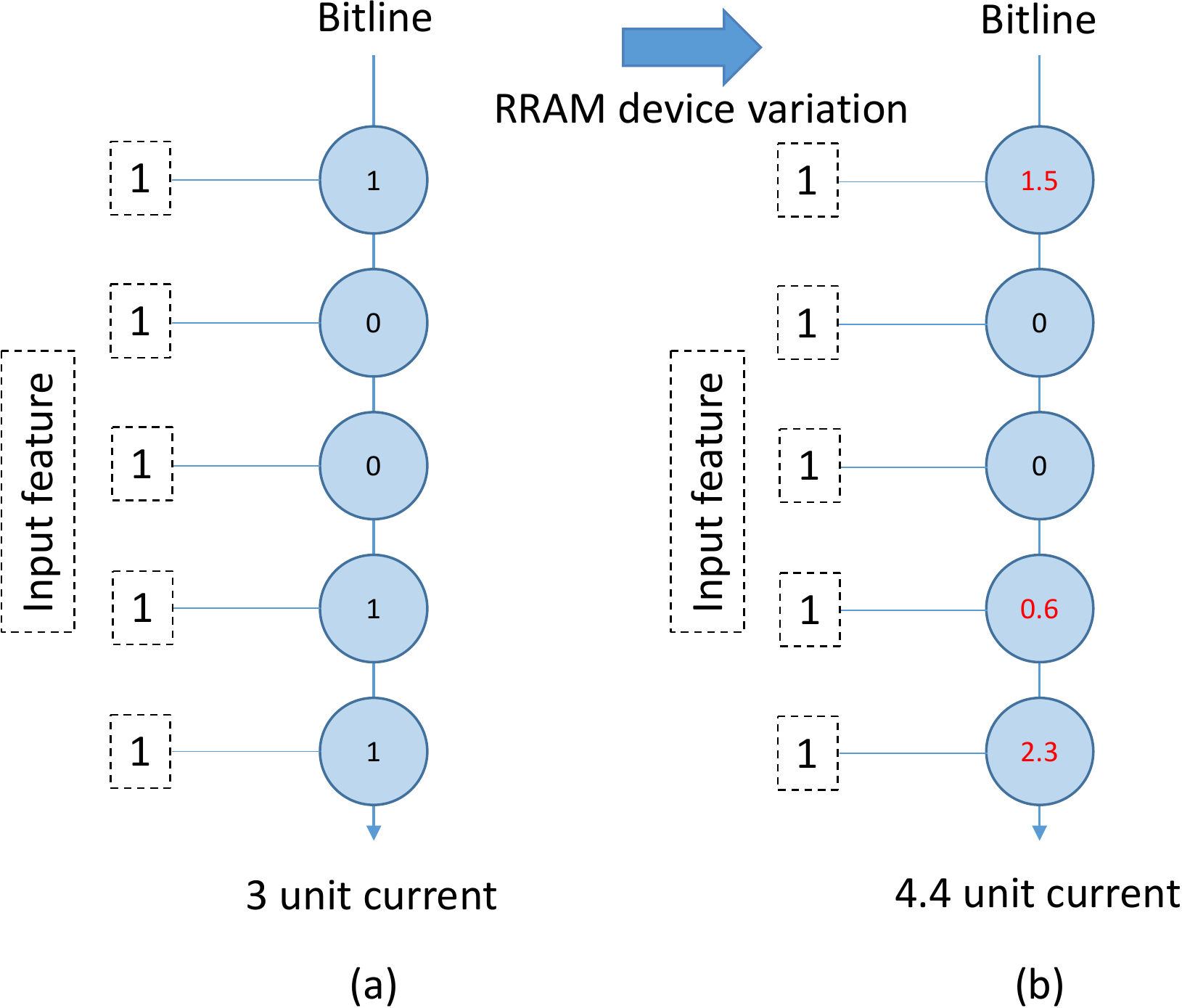}
  \caption[Example for RRAM cell device variation impact.]{{Example for the RRAM cell device variation.(a) The ideal current on the bit-line should be 3 units. (b) The actual current will become 4.4 units due to the variation.}}
  \label{fig:RRAM_device_variation_example}
\end{figure}
\newpage

\subsection{Device Variation}
RRAM device variation is inevitable because of the inherent stochasticity of RRAM and also because of the low word-line voltages chosen in this work that force the access transistor biased at the near sub-threshold region. Ideally, by multiplying weight one (low resistance state: LRS) and input one (Vdd), the current on the bit-line should equal exactly one unit regardless of the location. However, the measured resistance of the RRAM cells varies within the array and follows a log-normal distribution as shown in Fig.~\ref{fig:RRAM_device_variation_distribution}. Thus, the current on the bit-line deviates from the ideal value due to the cell resistance variation. 

Device variation leads to inaccurate accumulated current on the bit-line, i.e. wrong MAC results. Although the variation of binary RRAM cells is relatively small compared to the multilevel RRAM cell, it is not negligible. For instance, in Fig.~\ref{fig:RRAM_device_variation_example}, the ideal current accumulated on the bit-line should equal 3 units, but the actual current becomes 4.4 units due to the variation in individual cells. 

For high resistance state (HRS) cells, its device variation effect is less significant compared to the LRS cells due to its much higher resistance value and less contribution to the final accumulated current. In this paper, to simulate the device variation effect, we multiply a random mask generated by the log-normal distribution of the measured cell resistance before the convolution calculation.

Although the variation from a single RRAM cell looks substantial, the variation on the final bit-line current, i.e. the MAC value, remains tolerable because the variation distribution by summing a large number (1024) of RRAM cells becomes tighter than an individual cell according to the law of large numbers in statistics\cite{ITRICIM}.

\subsection{Nonlinearity of the bit-line Current Accumulation}
\begin{figure}[htb]
  \centering
  \includegraphics[height=!,width=0.9\linewidth,keepaspectratio=true]
  {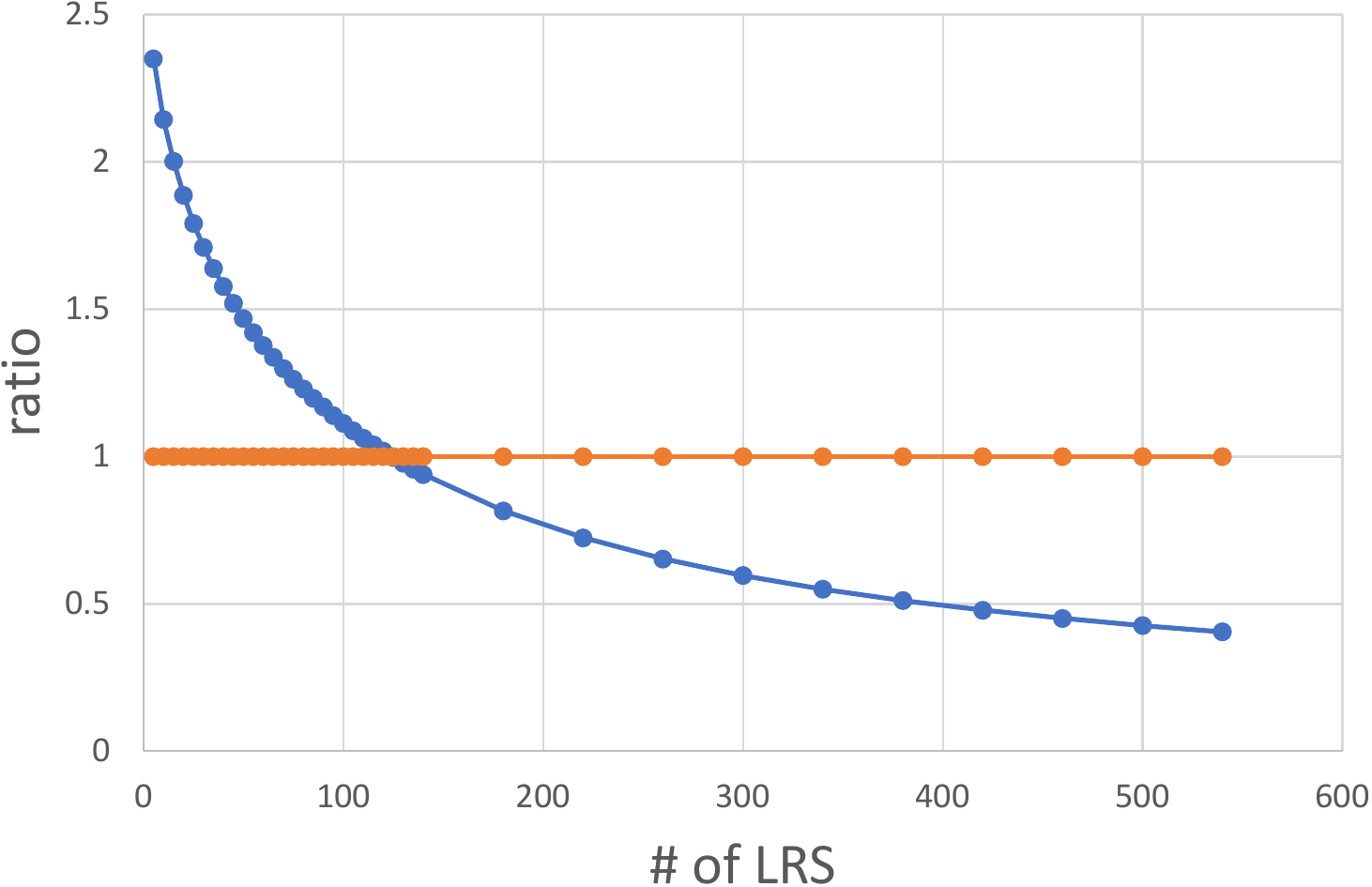}
  \caption[nonlinearity ratio of IRC macro.]{{Nonlinear ratio of the accumulated bit-line current vs activated LRS cells, where the orange line is the ideal one and the blue one is the nonideal one.}}
  \label{fig:non_linearity}
\end{figure}

Cell current on the bit-line depends on the cell resistance and bit-line clamping voltage. Ideally, the accumulated cell current on a bit-line is linearly proportional to the number of activated LRS cells and its ratio equals one as shown in  Fig.~\ref{fig:non_linearity}. However, this ratio is nonlinear and varies according to the number of activated LRS cells in the real chip because we do not use an operational amplifier to clamp the bit-line voltage fixed at 0.1V. The additional operational amplifier increases not only the area but also the power of the overall peripheral circuits. 

\begin{figure}[htb]
  \centering
  \includegraphics[height=!,width=1.0\linewidth,keepaspectratio=true]
  {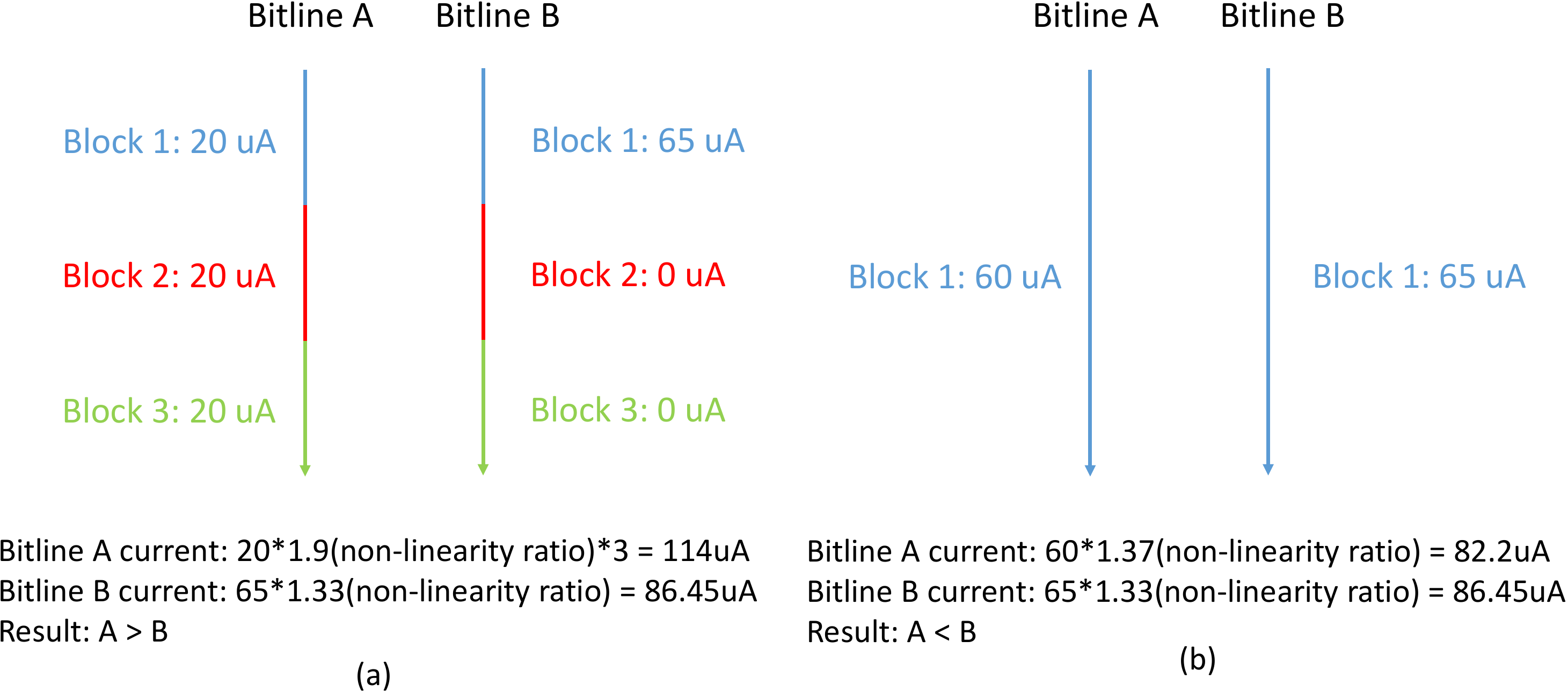}
  \caption[Example for nonlinearity impact.]{{\footnotesize(a) Accumulating the current partially and comparing them will get a wrong result. (b) Single accumulation and comparison can avoid the nonlinearity impact.}}
  \label{fig:non_linearity_example}
\end{figure}

This nonlinear effect significantly affects the accuracy of the MAC results if we only enable partial word-lines and accumulate these partial sums through an analog adder. This is because different values of partial sums possess different amounts of errors. By contrast, if the bit-line current of one complete convolution is accumulated in a single operation without using the partial sums, the nonlinearity does not directly change the final result if we only compare the relative current values of two bit-lines.
For example, in Fig.~\ref{fig:non_linearity_example}, if we split one convolution operation into three times and accumulate three bit-line currents externally, the total current becomes much larger than it should be. However, if we accumulate the total current in one single operation entirely in the crossbar array, the nonlinearity effect has little impact on the final result. 

We simulate the nonlinearity effect by multiplying the nonlinearity ratio with the current outputs from the crossbar array. The nonlinearity ratio based on the SPICE simulation of real circuits is fit using two piecewise high-degree polynomials as followings, where $p$ is the number of LRS cells.

\begin{equation}
 ratio =    \begin {cases}1.0286\times 10^{-8} \times p^4 - 3.79\times 10^{-6} \times p^3 \\+ 5.3\times 10^{-4} \times p^2 - 3.92\times 10^{-2} \times p + 2.5\\ if\: p <=140;
    \\1.8063\times 10^{-11} \times p^4 - 3.204\times 10^{-8} \times p^3 \\+ 2.2495\times 10^{-5} \times p^2 - 8.057\times 10^{-3} \times p + 1.707 \\ if\: p > 140; \end{cases}
\end{equation}

\subsection{Nonideal Peripheral Circuits}
The SA output is determined by the current difference of two bit-lines, i.e. the difference of the number of activated LRS cells on the two bit-lines. Its correctness depends on the current sensing capability of SA. The current accumulated on the bit-lines, either too large or too small, leads to incorrect results. Additionally, process variations may cause transistor mismatch and thus sensing variation in SAs. Therefore, in this work, we take into consideration the limited current sensing range and the sensing variation.

For the limited current sensing range, if the current is not within the designed range (in our IRC macro, the range is designed between 35$\mu A$ and 300 $\mu A$), a predictable output cannot be resolved. The maximum value of the accumulated current is limited by the bit-line width and its IR drop. The minimum value of the accumulated current is to ensure sufficient time to charge the parasitic capacitance on the bit-line, and thus the SA could determine the correct output result within the finite sensing time of interest.  In our simulation, this effect is considered by assigning random 0 or 1 SA output to those bit-lines with current outside of this sensing range. \par

\begin{figure}[htb]
  \centering
  \includegraphics[height=!,width=0.9\linewidth,keepaspectratio=true]
  {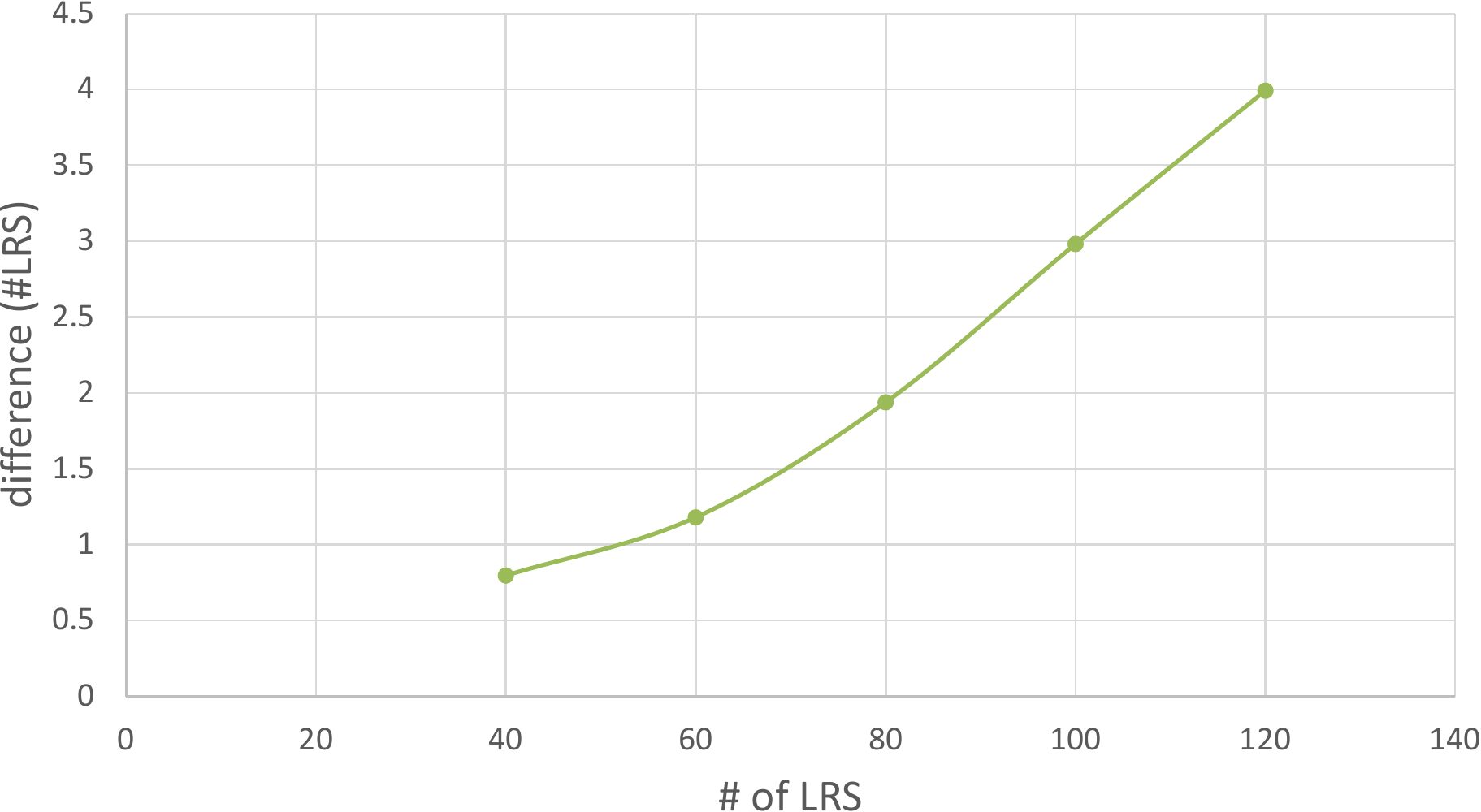}
  \caption[Offset value of SA sensing variation.]{{The required current difference in terms of the LRS cell number under different activated LRS cell numbers.}}
  \label{fig:SA_variation}
\end{figure}

For the sensing variation of SA, it can be modeled as an offset value for the input current of SA as shown in Fig.~\ref{fig:SA_variation} based on the Monte-Carlo simulation. Fig.~\ref{fig:SA_variation} shows the required current difference in terms of the number of LRS cells for correct detection. The more activated LRS cells are, the greater the current difference is required when considering the SA variation. Thus, the result of the binary activation could be wrong if the current values of the two bit-lines are close. We must ensure that the current differences of the two bit-lines are sufficiently large to distinguish the result successfully. We simulate the sensing variation of SA by adding a random offset based on the fitting curve of Fig. 9 by using a high-degree polynomial.\par

\subsection{IR drop}
IR voltage drop along the interconnect is inevitable due to the finite wire resistance, and it affects the final bit-line current because a less-than-expected voltage is dropped on the memory cell. In the IRC macro, the IR drop depends on the input pattern, weight mapping position, and parasitic resistance in the IRC crossbar array. Theoretically, the cells closer to the bit-line driver are less affected by the wire resistance. If more LRS cells are mapped farther away from the bit-line driver, the result of binary activation could be different from the expected due to the significant current reduction. 

\begin{figure}[htb]
  \centering
  \includegraphics[height=!,width=0.9\linewidth,keepaspectratio=true]
  {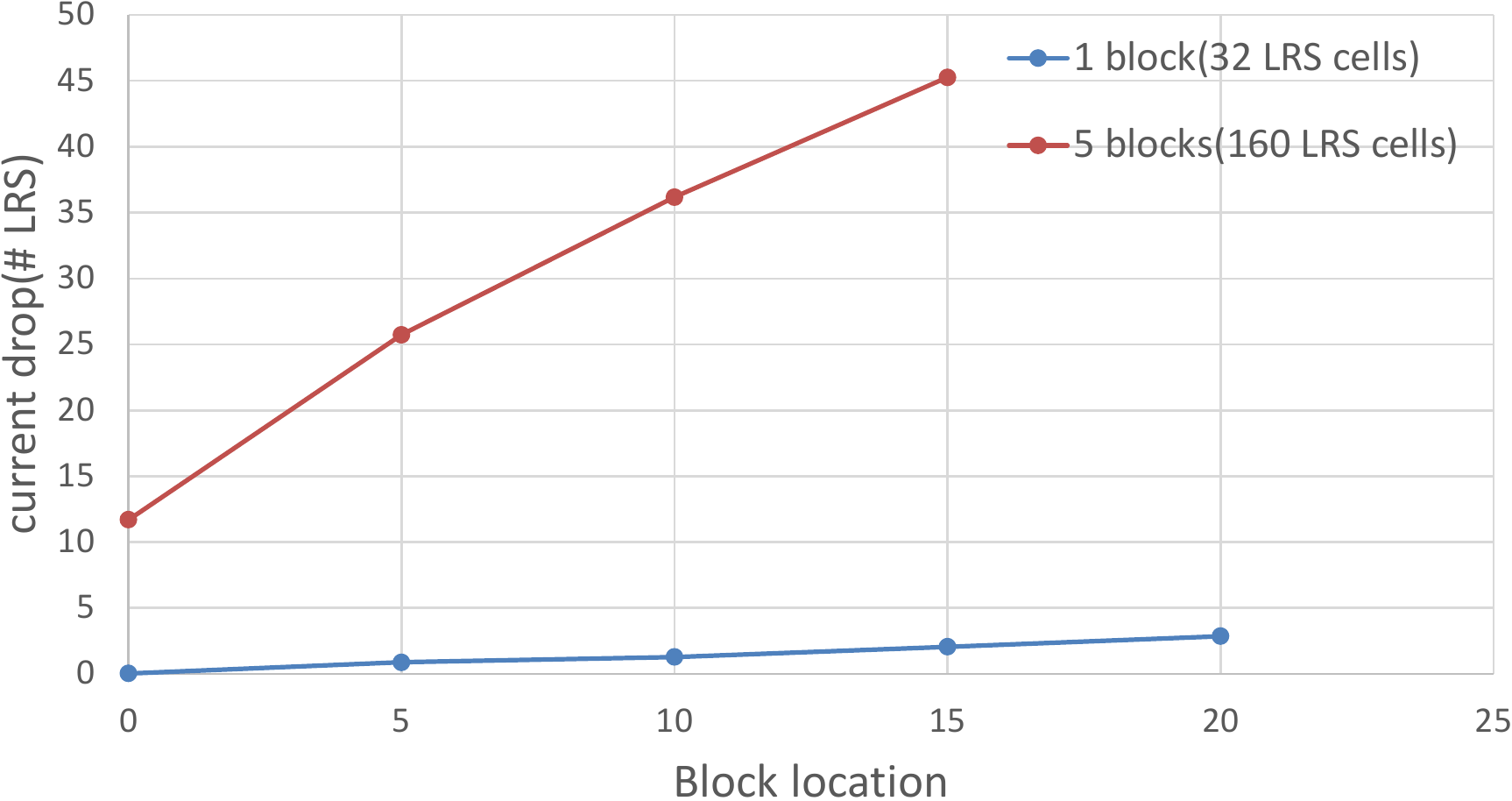}
  \caption[Relationship between LRS position and current drop.]{{Relationship between the LRS block location and current drop. The x-axis is the location of the block, where the smaller number means the block closer to the bit-line driver. The blue line is the current drop caused by 32 LRS cells in a single block.The red line is the current drop caused by 160 cells in 5 blocks from the position block 0 to block 4.}}
  \label{fig:wire_resistance_position}
\end{figure}

Fig.~\ref{fig:wire_resistance_position} shows the relationship between the weight position and the current reduction due to the IR drop. If the weights are mapped closer to the bit-line driver, the current reduction is smaller. The cost of simulating IR drop using SPICE is high as mentioned in \cite{he2019noise}. To simplify the simulation, we split the bit-line into several sub-blocks, and each sub-block contains 32 cells. The local IR drop effect among the cells in the same block is neglected, but the global IR drop effect is considered among the sub-blocks.  The calculated current of each cell depends on the input and weight pattern, and the result obtained using this method is close to that from a complete SPICE simulation. In most cases, the error is within 1\%.

\section{Object Detection with IRC}
\label{chapter:IRC_OD}

\subsection{Baseline Model and Design}
\begin{figure}[htb]
  \centering
  \includegraphics[height=!,width=1.0\linewidth,keepaspectratio=true]
  {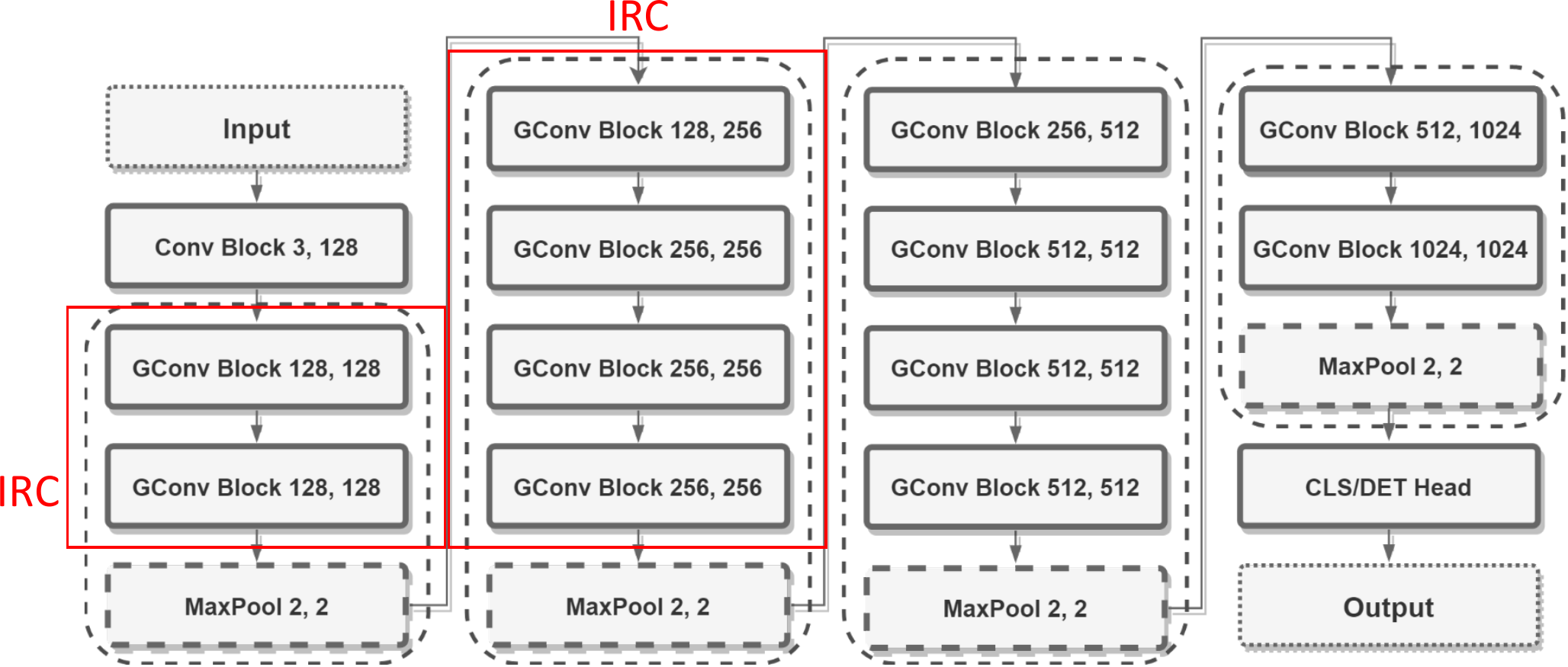}
  \caption[]{{ Overall model architecture for the IRC based object detection task, where the \textit{GConv Block m, n} is the binary group convolution with m input channels and n output channels. }}
  \label{fig:RRAM_original_model}
\end{figure}

The baseline BNN model for object detection is based on YOLOv2\cite{redmon2017yolo9000} but modified for IRC as shown in Fig.~\ref{fig:RRAM_original_model}, which uses binary group convolution with a group size of 60 and in-memory BN as the basic block. Only the layers inside the red boxes are implemented using the IRC macros because of the limitation of chip area. However, the simulation results are similar when all these group convolution layers are implemented using the IRC macros. The first and last layers are implemented using digital circuits as in other BNNs. For the model mapping, the baseline design uses a commonly used method by mapping one channel into one bit-line. In this model, because the kernel size is 3x3 and the channel size is 60, 636 (540 for convolution and 96 for BN bias) cells are required for one channel mapping. Therefore, a 1024x1024 array is adopted in our design. The convolution operation of one channel is split into several sub-blocks and accumulated sequentially in the analog domain for the final result as in Fig.~\ref{fig:non_linearity_example}. This is because the current limitation of the bit-line is set to 300$\mu A$ and too many activated cells exceeds this limit. The convolution result of a bit-line is compared to the reference bit-line by using a binary SA for the final activation output.

The baseline design has almost no tolerance to the nonideal effects as shown in the later result section due to three reasons. 
First, the baseline design uses binary weights for convolution and SAs for the current comparison between the convolution bit-lines and the reference bit-lines. Almost half of the cells on these bit-lines are LRS as a result of BNN. Therefore, the current will be significantly affected by the device variation and IR drop because having more LRS cells means the current value will be more unpredictable due to the variant resistance of the LRS cells as shown before.
Next, because the baseline model uses the BN layer and in-memory BN mapping, a portion of the cells on a bit-line will be used to map the bias value, implying that the bias value will also be significantly affected by the device variation and IR drop. However, in BNN, the BN precision is critical. Otherwise, this will result in a severe loss of accuracy.
Finally, because the convolution in a bit-line is split into sub-blocks to accumulate the current in the analog domain due to the bit-line current limitation, the nonlinearity effect is significantly increased  as shown in Section III. These problems will be solved by the following proposed approaches.

\subsection{Proposed methods}
The proposed hardware robust IRC design for object detection is shown below. The object detection, which primarily consists of classification and regression, was chosen to demonstrate how nonideal effects affect this more complex task. We propose ternary weight mapping, BN removal, extra bias on the bit-line, and lower wordline voltage to recover the loss due to nonideal effects, as discussed below.

\begin{figure}[htb]
  \centering
\includegraphics[height=!,width=0.5\textwidth,keepaspectratio=true]
  {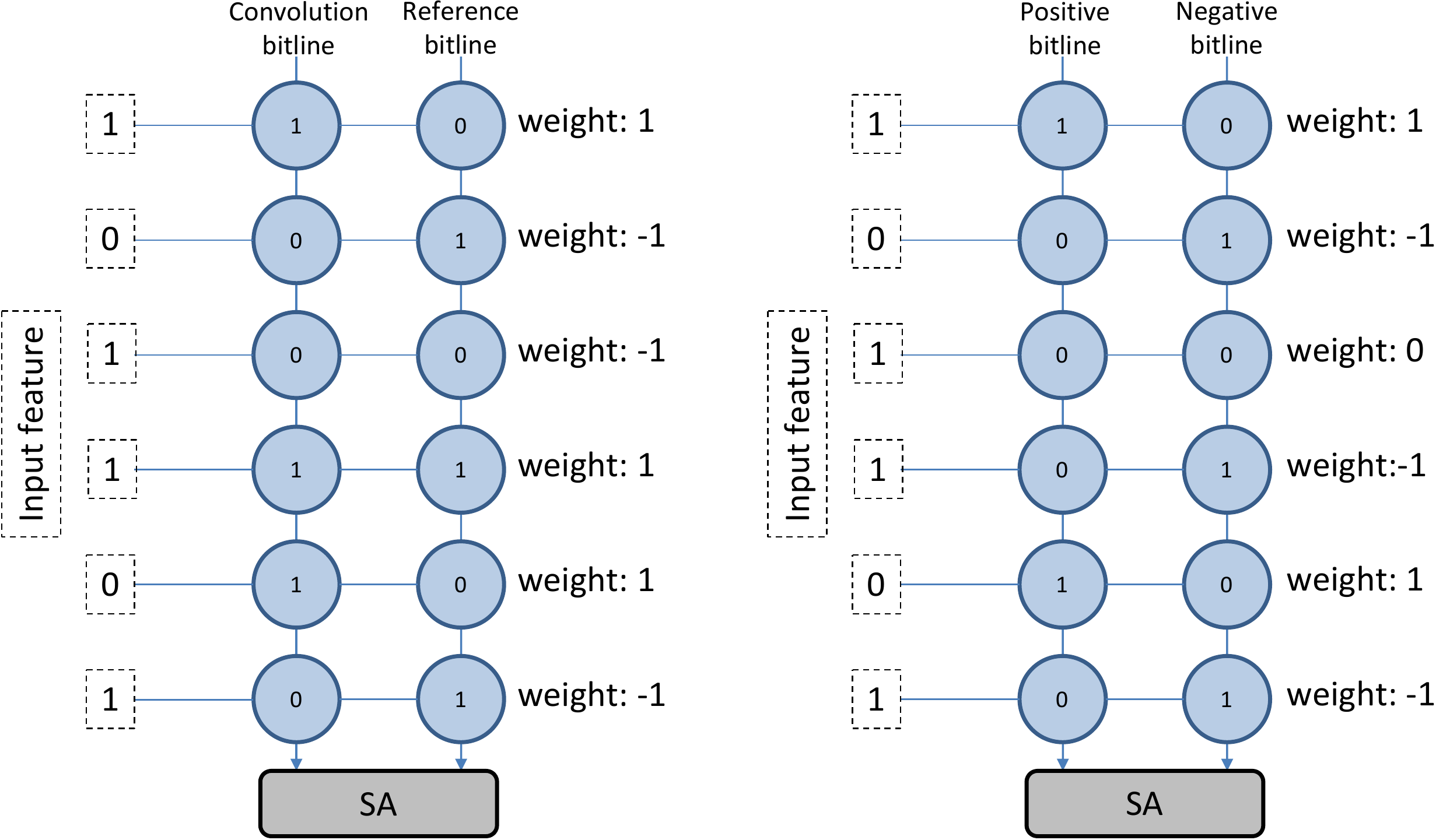}
  \caption[Ternary weight mapping for IRC macro.]{{Weight mapping for the IRC crossbar array, where 1 means LRS, and 0 means HRS. (a) (left) binary weight mapping: (1, 0) at the convolution bit-line means weight values 1, and -1, respectively, and the reference bit-line has even distributed 0 and 1. (b) (right) ternary weight mapping: (1, 0), (0, 1), and (0, 0) pairs at the same wordline means weight values 1, -1, and 0, respectively. }}
  \label{fig:ternary_mapping}
\end{figure}

\subsubsection{Ternary weight mapping}
The baseline model maps the binary weight to one bit-line and one common reference bit-line. However, if the reference bit-line is incorrect, all results of the convolution bit-line will have an extra input dependent offset. To solve this problem and improve the accuracy, we use ternary weights (0, +/-1) rather than binary weights (+/-1). The ternary weight mapping will use one positive bit-line (G+) and negative bit-line (G-) as shown in Fig.~\ref{fig:IRC_macro} for current comparison. For a +1 weight, the positive bit-line cells will be LRS, and the negative bit-line cells will be HRS. For a -1 weight, the positive bit-line cells will be HRS, and negative bit-line cells will be LRS. For a 0 weight, both positive and negative bit-line cells will be HRS, as shown in Fig.~\ref{fig:ternary_mapping}.

We can further regulate the ternary weight distribution to 20\%, 60\%, and 20\% for -1, 0, and 1, respectively, in each group of filters for better hardware robustness and lower power usage. This can increase the ratio of HRS cells of each bit-line, reducing the impact of device variation and IR drop and assisting in keeping the current within the expected range for lower power and better linearity. With this regularization, the cell distribution in this model will be 20\% for LRS, and 80\% for HRS. \par

\subsubsection{BN removal}
\begin{figure}[htb]
  \centering
  \includegraphics[height=!,width=1.0\linewidth,keepaspectratio=true]
  {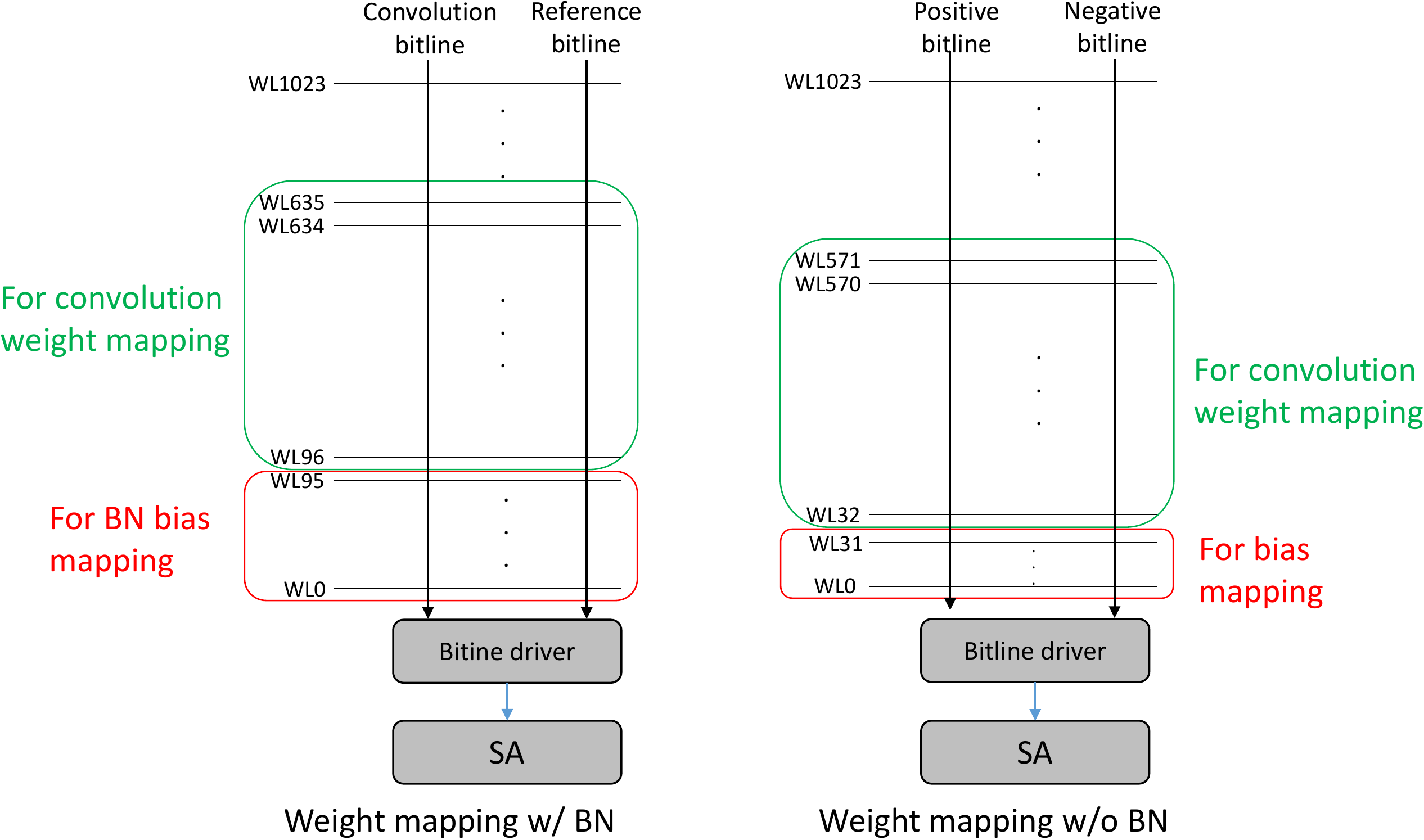}
  \caption[Mapping position for model without.]{{(a) In the original model, the 96 cells closest to the bit-line driver are used for BN mapping for better BN precision. (b) In the proposed model, only 32 cells are needed for extra bias mapping.}}
  \label{fig:model_without_BN}
\end{figure}

We remove BN from the IRC-implemented layers. Without BN, the number of LRS cells in a bit-line will be reduced due to BN mapping, which means that not only will the influence of device variation be reduced, but also the current drop caused by IR drop will be reduced. With the BN removed, the convolution cells can be placed closer to the bit-line driver, as shown in Fig.~\ref{fig:model_without_BN}.

\begin{figure}[htb]
  \centering
  \includegraphics[height=!,width=1.0\linewidth,keepaspectratio=true]
  {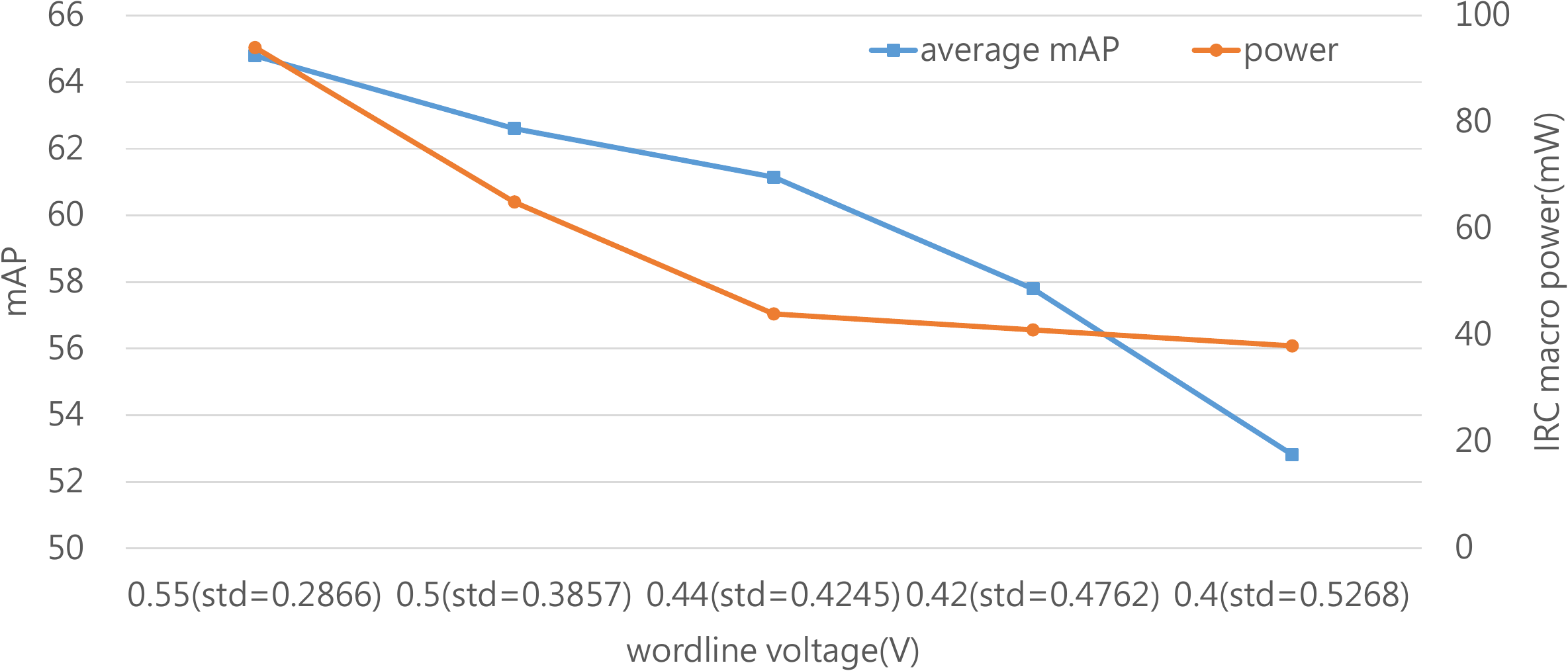}
  \caption[]{{Relationship between the wordline voltage, power of IRC macro, and mAP results, where the number $A(std=B)$ in the x-axis denotes the wordline voltage $A$ and device variation $B$.}}
  \label{fig:wl_voltage}
\end{figure}

\subsubsection{Lower wordline voltage}
Because of the maximum current limitation, the baseline model cannot activate all wordlines. To activate all wordlines under this constraint, we reduce the wordline voltage of the IRC macro, which can reduce the current accumulated on a bit-line. As a result, we can accumulate all cell current in a single operation. Although this raises the standard deviation of the device variation distribution, it is justified because it reduces the impact of nonlinearity. Fig.~\ref{fig:wl_voltage} shows the relationship between wordline voltage, power of IRC macro and device variation. The condition with 0.44(std=0.4245) in the figure is the kink point for power consumption and has mAP greater than 60. Thus, we choose this point for a better balance between power and accuracy. \par

\subsubsection{Extra bias for limited sensing range}
The methods mentioned above could reduce the influence of device variation, nonlinearity, and IR drop. However, the limited sensing range remains a challenge. For the upper limit of the current range, our ternary weight model can ensure that no such situation will occur due to only 20\% of LRS cells in the channel. The major concern is the lower limit. If the current difference of the two bit-lines is less than the lower limit, it cannot be detected at all. This lower limit can be regarded as an offset. This situation is the most common case in the convolution due to its quite symmetrical data distribution as shown in Table~\ref{table:extra_bias}. In some layers, this will result in 29.28\% of bit-line current detection failure. One solution is to add an extra bias to compensate this offset. 

The choice of the bias value is the trade-off between the effects of sensing variation of SA and the limited current sensing range. To balance these two effects, we choose the bias value in a layerwise way, as shown in Table~\ref{table:extra_bias}. Without the extra bias mapping, a high percentage of current difference will be less than the lower bound of the sensible value, resulting a significant performance drop. We could balance the influence of these two nonideal effects with a suitable bias and reduce the failure case to less than 3\% to minimize the performance impact. \par

\begin{table*}[htb]
\centering
\caption[Choice of extra bias value to balance the impact between the effects of sensing variation of SA and limited current sensing range.]{The ratio of values changed by sensing variation of SA and nonsensible. With the extra bias mapping, the impact of these two effects is more balanced.}
\label{table:extra_bias}
\begin{adjustbox}{width=1\textwidth}
\small
\begin{tabular}{@{}cccccccc@{}}
\toprule
SA errors                             &                                 & Layer2\_0 & Layer2\_1 & Layer3\_0 & Layer3\_1 & Layer3\_2 & Layer3\_3 \\ \midrule
sensing variation                     & \multirow{2}{*}{w/o extra bias} & 2.35\%    & 1.17\%    & 1.53\%    & 2.31\%    & 2.21\%    & 1.24\%    \\
bit-line current \textless lower bound &                                 & 9.08\%    & 2.17\%    & 29.28\%   & 2.32\%    & 7.69\%    & 1.18\%    \\ \midrule
sensing variation                     & \multirow{2}{*}{w/ extra bias}  & 2.87\%    & 1.36\%    & 2.63\%    & 2.39\%    & 2.38\%    & 1.33\%    \\
bit-line current \textless lower bound &                                 & 2.74\%    & 0.52\%    & 2.93\%    & 0.94\%    & 2.52\%    & 0.25\%    \\ \bottomrule
\end{tabular}
\end{adjustbox}
\end{table*}

\section{Experimental Results}
\subsection{Settings}
The baseline and proposed models are implemented using Pytorch and evaluated on the IVS 3cls~\cite{tsai20202020} dataset for object detection. This dataset contains 10,000 training images and 1,000 test images with three categories of objects: vehicles, bike, and pedestrian. The images have a resolution of 1920$\times$1080 and are rescaled to 1024$\times$576 as our input. For model training, we use the AdamW~\cite{loshchilov2017decoupled} optimizer with 10$^{-3}$ weight decay. The learning rate is warmed up from 10$^{-5}$ to 10$^{-4}$ at the first five epochs, and finally reduced to 10$^{-5}$ and 10$^{-6}$ at the 80th and 110th epoch, respectively. The mini-batch size is set to 6. The nonideal effects are modeled as mentioned before and added to the inference phase with 10 random seeds to simulate their randomness.

\subsection{Models With or Without BN}
\begin{figure}[htb]
  \centering
  \includegraphics[height=!,width=1.0\linewidth,keepaspectratio=true]
  {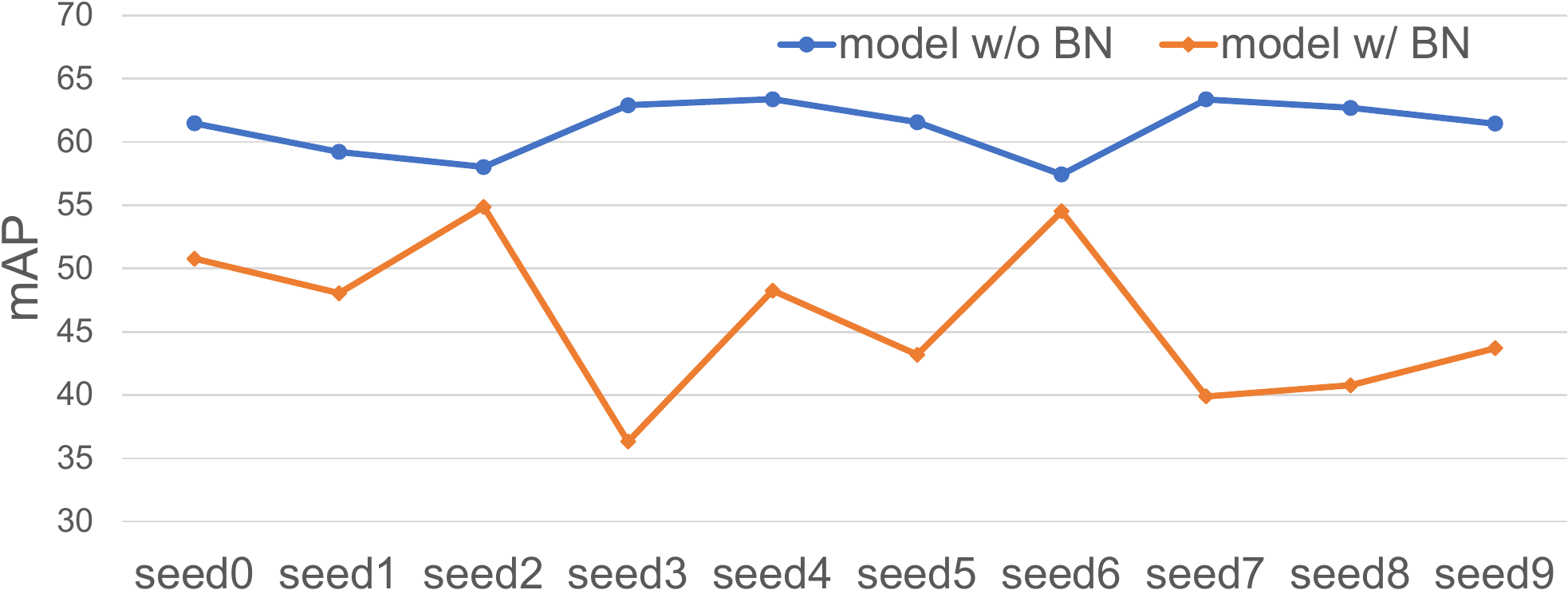}
  \caption[]{mAP results considering device variation for models with and without BN.}
  \label{fig:device_variation_with_BN}
\end{figure}

Fig.~\ref{fig:device_variation_with_BN} shows the evaluation result of the device variation on BN with 10 different variation seeds. The mAP for the model with and without BN is 61.15\%$\pm$2.19 and 46.03\%$\pm$6.28, respectively. When considering the device variation, the average mAP results for the model with BN show a significant degradation and larger standard deviation. Device variation has a great impact on BN precision due to in-memory BN, resulting in a significant model accuracy drop. \par

\begin{figure}[htb]
  \centering
  \includegraphics[height=!,width=1.0\linewidth,keepaspectratio=true]
  {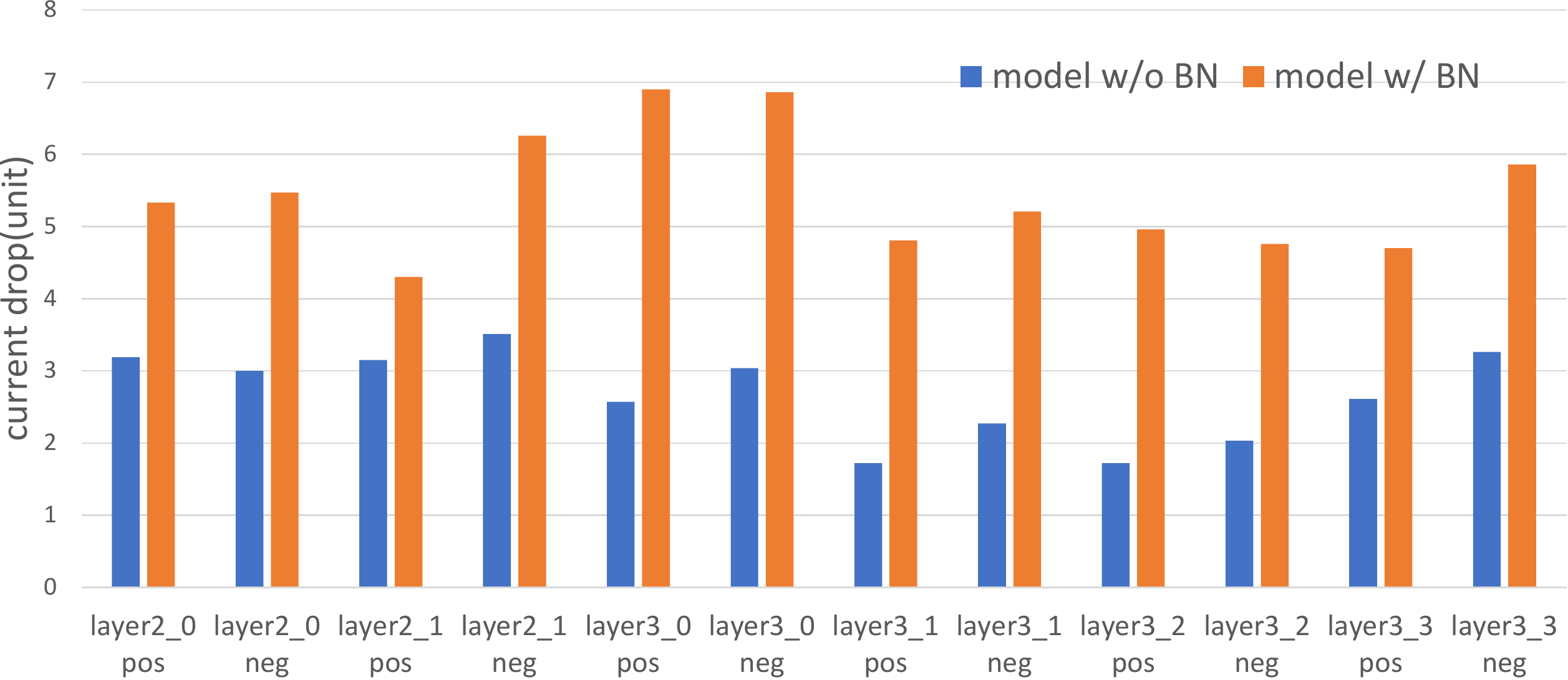}
  \caption[Current drop value comparison between model with and without BN.]{Current drop comparison between models with and without BN layers, where pos is the positive bit-lines and neg is the negative bit-lines.}
  \label{fig:current_drop_BN}
\end{figure}

We also examine the IR drop issue caused by BN bias mapping. Fig.~\ref{fig:current_drop_BN} shows the average current drop in different layers. The average current drop for the model with BN is roughly twice that of the model without BN, implying that the BN layer would exacerbate the IR drop problem and further degrade performance.

\subsection{Simulation Result of the Proposed Model}
Table~\ref{table:OD_result} shows the simulation results and comparisons between the original and proposed models. Our model becomes more resistant to the nonideal effects with the proposed method. As shown in the table, the device variation has a significant impact on the original model (32.8\% mAP drop) but only a minor impact on the proposed model (4.3\% mAP drop). Furthermore, the original model will fail completely due to the nonlinearity, whereas the new model is robust and maintains nearly the same performance. Nonideal peripheral circuits and IR drop have a lower impact than device variation, but they still cause an additional 1.9\% and 0.9\% mAP drop, respectively. The mAP results for these two effects in the original model are messed up. Except for IR drop, the degradation of the proposed model due to these effects can be restored with retraining. The IR drop effect is assessed after retraining rather than during retraining since it takes too long to retrain. If IR drop is included in the retraining process, as it is in other works, the outcome could be improved.

\begin{table}[h]
\centering
\caption[mAP results for original model and proposed model.]{mAP results for the original and proposed models.}
\label{table:OD_result}
\begin{tabular}{@{}ccccccccc@{}}
\toprule
Ideal                &\checkmark      &           &           &           &           &                &                \\ \midrule
Device variation     &                &\checkmark &\checkmark &\checkmark &\checkmark &\checkmark      &\checkmark      \\
nonlinearity        &                &           &\checkmark &\checkmark &\checkmark &\checkmark      &\checkmark      \\
nonideal peri. ckts&                &           &           &\checkmark &\checkmark &\checkmark      &\checkmark      \\
IR drop              &                &           &           &           &\checkmark &                &\checkmark      \\
Retraining           &                &           &           &           &           &\checkmark      &\checkmark      \\ \midrule
Original        & 65.1          & 32.3     & 0.0       & -         & -         & -              & -              \\ \midrule
proposed (\textit{avg})       & \textbf{65.4} & 61.1     & 61.1     & 59.2     & 58.3     & \textbf{63.2} & \textbf{61.5} \\ 
\textit{std deviation}     &                  & 2.2  & 2.2  & 2.0  & 2.2  & 1.2  & 2.0  \\ \bottomrule 
\end{tabular}
\end{table}

\begin{figure*}[htb]
\centering
\subfigure[]{
\label{fig:OD_demo_ideal}
\adjustbox{trim={.0\width} {.5\height} {0.0\width} {.0\height}, clip}%
{\includegraphics[width=0.45\linewidth,keepaspectratio=true]{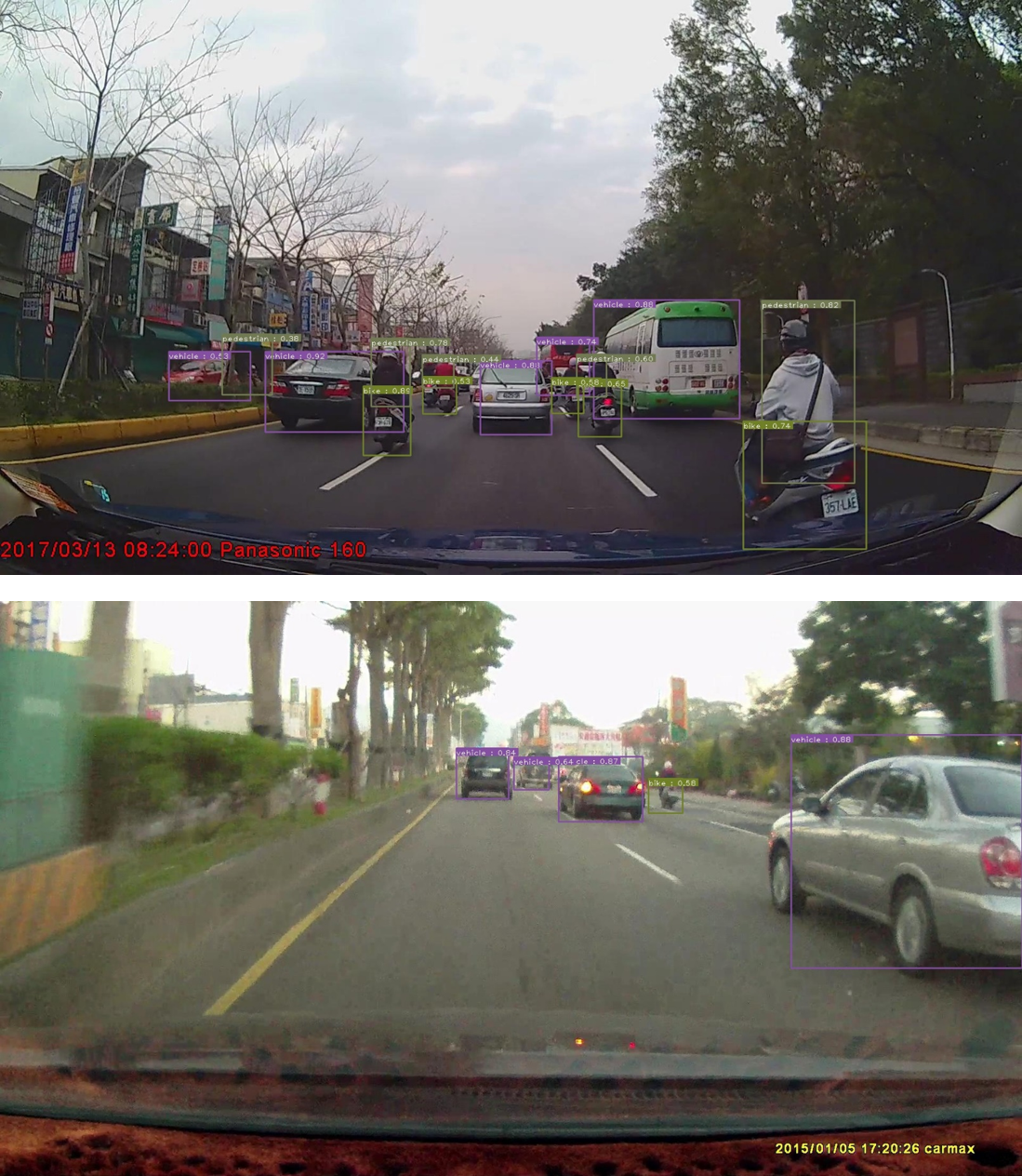}}}
\subfigure[]{
\label{fig:OD_demo_non_ideal}
\adjustbox{trim={.0\width} {.5\height} {0.0\width} {.0\height}, clip}%
{\includegraphics[width=0.45\linewidth,keepaspectratio=true]{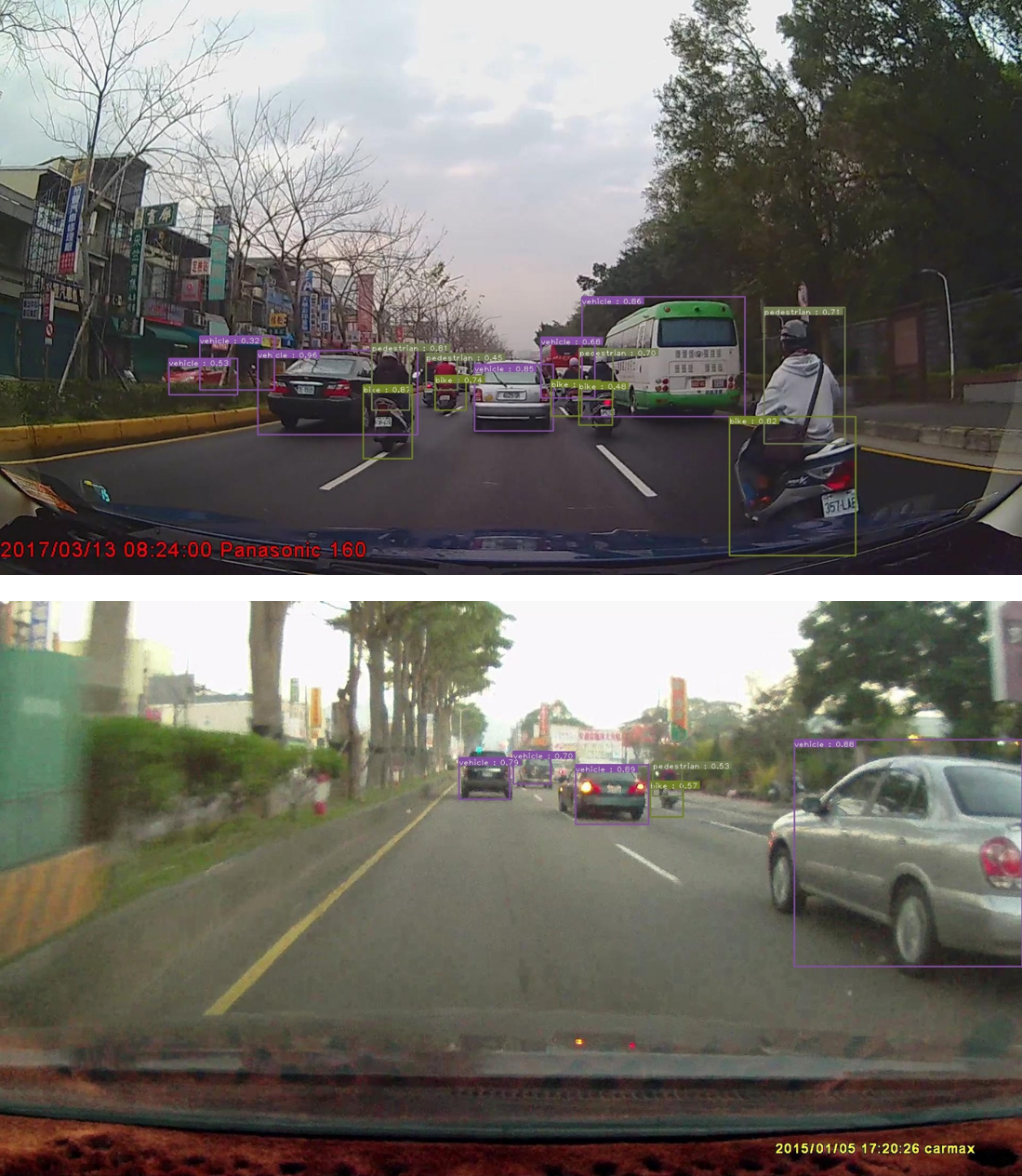}}}
\centering
\caption{Demo results of (a) the ideal model and (b) the proposed model with the proposed approach under nonideal effects.}
\label{fig:OD_demo}
\end{figure*}

Fig.~\ref{fig:OD_demo} shows the test images for different models. In this test image, the result is almost the same between the ideal model (Fig.~\ref{fig:OD_demo_ideal}) and the proposed model considering the nonideal effects (Fig.~\ref{fig:OD_demo_non_ideal}).

\begin{figure}[htb]
  \centering
  \includegraphics[height=!,width=1.0\linewidth,keepaspectratio=true]
  {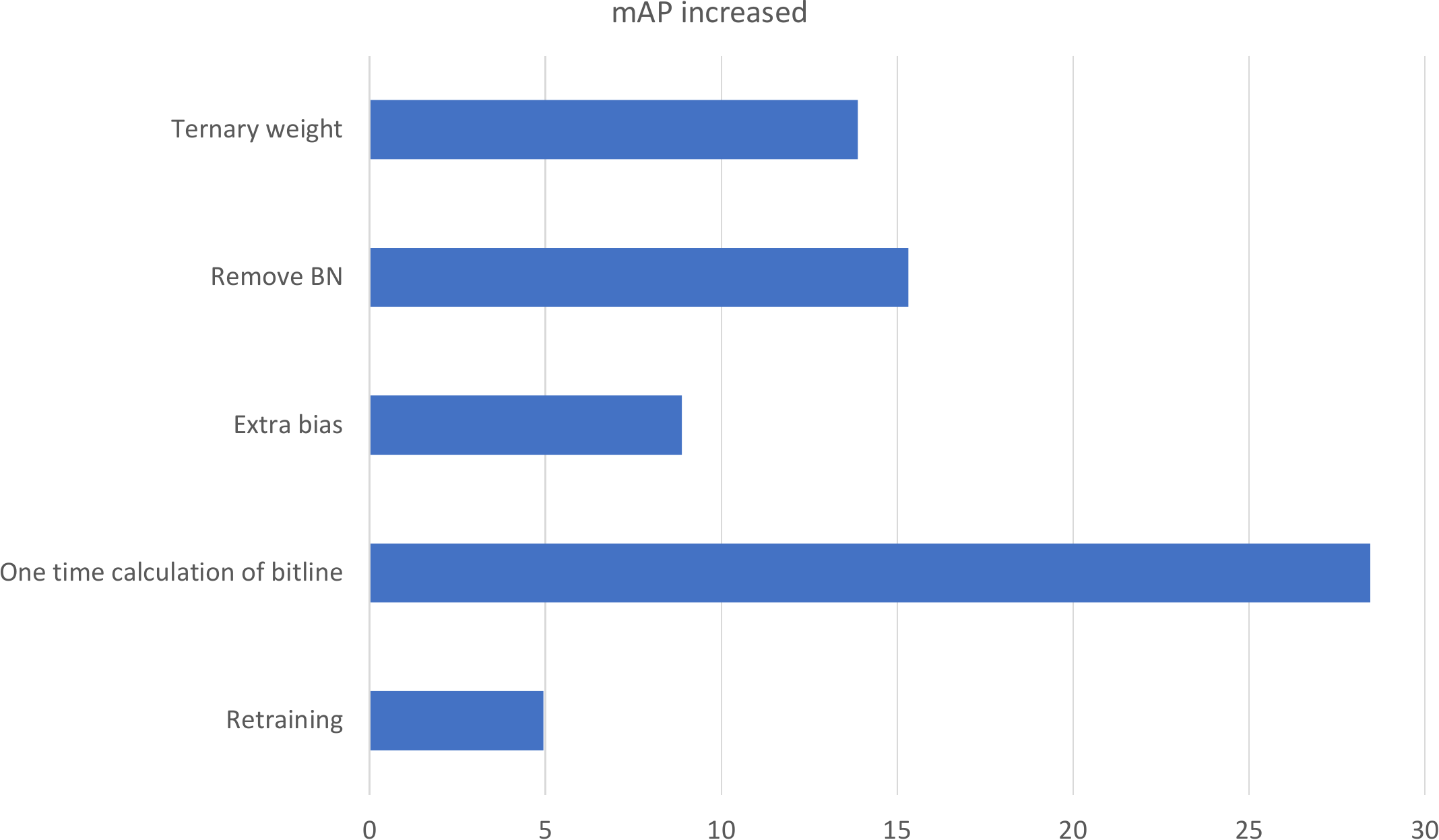}
  \caption[]{{mAP increase due to different methods.}}
  \label{fig:mAP_increased}
\end{figure}

To better understand the improvement of our approach to different nonideal effects, Table~\ref{table:methods_effects} shows the relationship between different methods and the nonideal effects. In which, the +/- symbol is positive or negative to reduce the nonideal effects. Fig.~\ref{fig:mAP_increased} shows the improvement of our method on the mAP results. The lower wordline voltage enables a single calculation of a bit-line and solves the nonlinearity problem well, which contributes most of the mAP improvement. The ternary weight model without BN could mitigate the impact of device variation and IR drop, which contributes the second largest improvement. The extra bias is used to balance the sensing variation and current limitation of SA circuits, which is also a good approach. Finally, retraining as in other works can also improve the final result slightly.  In summary, making the IRC macro robust through joint hardware and software optimization is the first step to get a good result. \par

\begin{table}[htb]
\centering
\caption{Pros and cons of our approaches on the nonideal effects of the IRC macro.}
\label{table:methods_effects}
\begin{tabular}{|l|l|l|l|l|}
\hline
               & Ternary& Remove BN & Extra& Lower WL\\ 
               & weight &           & bias & voltage \\ \hline
Cell variation & +              & +         &            & -              \\ \hline
nonlinearity  &                &           &            & +              \\ \hline
SA variation   & +              & +         & -          &                \\ \hline
Sensing range  &                &           & +          &                \\ \hline
IR drop        & +              & +         & -          &                \\ \hline
\end{tabular}
\end{table}

To determine the limitation of our robust model, we also simulate the mAP results with larger device variation and sensing variation of SA. The target is to maintain the mAP result above 60 and find the largest acceptable variation. Table~\ref{table:variation_limitation} shows the results. The largest acceptable standard deviation for device variation is about 0.43 to 0.44, which is quite close to our choice in previous experiments. For sensing variation of SA, our model can tolerate a more nonideal SA circuit. In our experiment, the misjudgment of the current range can be 2 to 3 units larger compared with Fig.~\ref{fig:SA_variation}.

\begin{table}[htb]
\centering
\caption[The tolerance limitation of our robust model.]{The tolerance limitation of our robust model. Sensing variation of SA + 1 means the variation is 1 unit larger than that in Fig.~\ref{fig:SA_variation}.}
\label{table:variation_limitation}
\begin{tabular}{@{}cccc@{}}
\toprule
Device std variation   & 0.44 & 0.43 & 0.42 \\
mAP & 61.48                         & 58.25                         & 55.58                         \\ \midrule
Sensing variation    &  + 1   &  + 2   &  + 3   \\
mAP & 64.81                         & 63.49                         & 59.60                         \\ \bottomrule
\end{tabular}
\end{table}

\begin{table*}[htb]
\centering
\caption[Nonideal effects comparison with other works.]{Comparisons with other works under different nonideal effects.}
\label{table:non_ideal_comparison}
\begin{adjustbox}{width=1\textwidth}
\begin{tabular}{@{}cccccccc@{}}
\toprule
       & Device Variation & Non-liearity & IR drop & Peripheral & Stuck-at fault & Accuracy drop & Task(dataset)                      \\
NIA\cite{he2019noise}  &                  &              & \checkmark       &            & \checkmark     & 1(\%)         & classification(MNIST)      \\
IR-QNN\cite{fouda2020ir} & \checkmark       & \checkmark   & \checkmark       &            & \checkmark     & 1.45(\%)      & classification(CIFAR-10)   \\
SCN\cite{lee2020learning}    &                  &              & \checkmark       &            &                & 4.53(\%)      & classification(CIFAR-10)   \\ \midrule
Ours   & \checkmark       & \checkmark   & \checkmark       & \checkmark &                & 3.85(\%mAP)     & object detection(IVS 3cls) \\ \bottomrule
\end{tabular}
\end{adjustbox}
\end{table*}

\subsection{Comparisons with Other Approaches}
Table~\ref{table:non_ideal_comparison} shows the results and comparisons of various works. Compared to other works, we have covered more nonideal effects to ensure reliability. In contrast to previous works, these nonideal effects are based on simulations as well as chip measurement.  Furthermore, our study is the first to analyze and simulate these effects on a more complex  object detection task while maintaining nearly identical performance.

\section{Conclusion}
This paper presents a hardware robust in-RRAM-computing design for object detection. We first analyze the possible nonideal effects for a large fully parallel operated IRC macro using simulations and real chip measurements, and we discover that the traditional retraining approach is incapable of resolving these effects well. Instead, we jointly optimize the IRC macro robustness through hardware and software. We lower the wordline voltage to enable a complete convolution in one operation to minimize the nonlinear addition. Furthermore, we use ternary weight mapping and remove BN to increase the tolerance to device variation, SA variation, and IR drop. An extra bias is added to overcome the limited range of current sensing. The proposed approach can keep the mAP loss below 4\% when directly executing a complex object detection model under hardware nonlinearity. The proposed approach can be applied to other types of in-memory-computing as well. Furthermore, extension to multibit networks will be our future work for more aggressive applications.

\bibliographystyle{IEEEtran}

\bibliography{IEEEabrv, bib/thesis}

\begin{IEEEbiography}[{\includegraphics[width=1in,height=1.25in,clip,keepaspectratio]{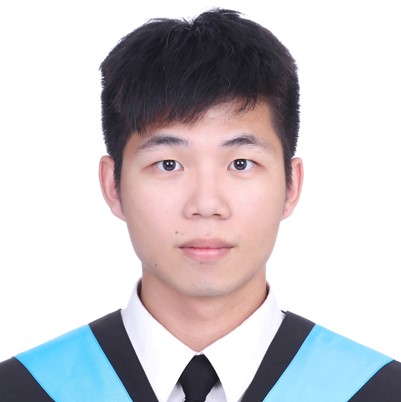}}]{Yu-Hsiang Chiang}
 received the M.S. degree in electronics engineering from the National Yang Ming Chiao Tung University, Hsinchu, Taiwan, in 2021. He is currently working in the Mediatek, Hsinchu, Taiwan. His research interest includes in-memory computing architecture design and VLSI design.

\end{IEEEbiography}

\begin{IEEEbiography}[{\includegraphics[width=1in,height=1.25in,clip,keepaspectratio]{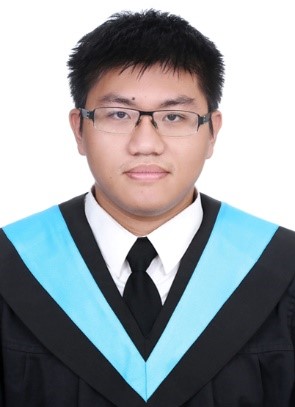}}]{Cheng-En Ni}
received the M.S. degree in electronics engineering from the National Yang Ming Chiao Tung University, Hsinchu, Taiwan, in 2021. He is currently working in the Novatek, Hsinchu, Taiwan. His research interest includes VLSI design and deep learning.

\end{IEEEbiography}
\begin{IEEEbiography}[{\includegraphics[width=1in,height=1.25in,clip,keepaspectratio]{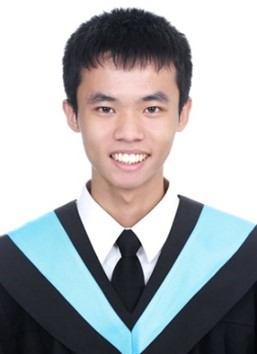}}]{Yun Sung}
received the B.S. degree in undergraduate honor program of nano science and engineering from the National Chiao Tung University, Hsinchu, Taiwan, in 2019. He is currently pursuing the M.S. degree with the institute of electronics, National Yang Ming Chiao Tung University, Hsinchu, Taiwan. His research interests include VLSI and in-memory computing architecture design.

\end{IEEEbiography}
\begin{IEEEbiography}[{\includegraphics[width=1in,height=1.25in,clip,keepaspectratio]{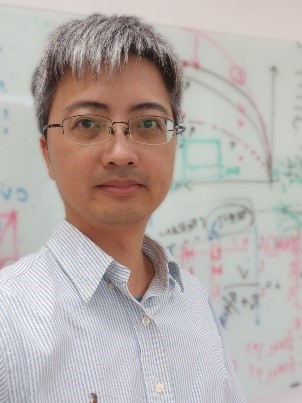}}]{Tuo-Hung Hou}
	(S’05–M’08–SM’14)
	received his Ph.D. degree in electrical and computer engineering from Cornell University in 2008. In 2000, he joined Taiwan Semiconductor Manufacturing Company (TSMC). Since 2008, he joined the Department of Electronics Engineering, National Chiao Tung University (NCTU) (as National Yang Ming Chiao Tung University (NYCU) in 2021), where he is currently a Distinguished Professor and Associate Vice President for Research \& Development. His research interests include the emerging nonvolatile memory for embedded and high-density data storage, electronic synaptic device and neuromorphic computing systems, and heterogeneous integration of silicon electronics with low-dimensional nanomaterials.
\end{IEEEbiography}
\begin{IEEEbiography}[{\includegraphics[width=1in,height=1.25in,clip,keepaspectratio]{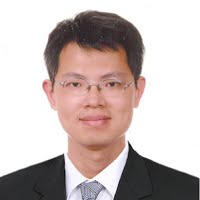}}]{Tian-Sheuan Chang}
	(S’93–M’06–SM’07)
	received the B.S., M.S., and Ph.D. degrees in electronic engineering from National Chiao-Tung University (NCTU), Hsinchu, Taiwan, in 1993, 1995, and 1999, respectively. 
	
	From 2000 to 2004, he was a Deputy Manager with Global Unichip Corporation, Hsinchu, Taiwan. In 2004, he joined the Department of Electronics Engineering, NCTU (as National Yang Ming Chiao Tung University (NYCU) in 2021), where he is currently a Professor. In 2009, he was a visiting scholar in IMEC, Belgium. His current research interests include system-on-a-chip design, VLSI signal processing, and computer architecture.
	
	Dr. Chang has received the Excellent Young Electrical Engineer from Chinese Institute of Electrical Engineering in 2007, and the Outstanding Young Scholar from Taiwan IC Design Society in 2010. He has been actively involved in many international conferences as an organizing committee or technical program committee member.
\end{IEEEbiography}
\begin{IEEEbiography}[{\includegraphics[width=1in,height=1.25in,clip,keepaspectratio]{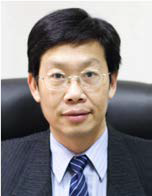}}]{Shyh-Jye Jou}
received his B. S. degree in electrical engineering from National Chen Kung University in 1982, and M. S. and Ph.D. degrees in electronics from National Chiao Tung University in 1984 and 1988, respectively. He joined Electrical Engineering Department of National Central University, Chung-Li, Taiwan, from 1990 to 2004 and became a professor in 1997. Since 2004, he has been a professor at the Electronics Engineering Department of National Chiao Tung University (as National Yang Ming Chiao Tung University (NYCU) in 2021) and served as the Chairman from 2006 to 2009. He was appointed as Director General of Ministry of Science and Technology in charge of Science Education and International Cooperation department from Jan. 2016 to Dec. 2017. He was a visiting research professor in the Coordinated Science Laboratory at University of Illinois, Urbana Champaign during 1993-1994 and 2010 academic years. In the summer of 2001, he was a visiting research consultant in the Communication Circuits and Systems Research Laboratory of Agere Systems, USA. He received Outstanding Engineering Professor Award, Chinese Institute of Engineers and Chinese Institute of Electrical Engineering at 2011 and 2013, respectively. He served as 2006 Chapter Chair of IEEE Circuits and Systems Society Taipei Chapter, 2009-2010 Distinguished Lecturer of CAS society, Guest Editor of IEEE Journal of Solid State Circuits, Nov. 2008, and was Track Chair, 2011-2013 Nanoelectronics and Gigascale Systems. He also serves as conference chairs, TPC chairs in many IEEE conferences such as Conference Chair of IEEE VLSI-DAT and International Workshop on Memory Technology, Design, and Testing. He also served as Technical Program Chair or Co-Chair in IEEE VLSI-DAT, International IEEE Asian Solid-State Circuit Conference, IEEE Biomedical Circuits and Systems. He has published more than 100 IEEE journal and conference papers. His research interests include design and analysis of high speed, low power mixed-signal integrated circuits, communication and Bio-Electronics integrated circuits and systems.

\end{IEEEbiography}
\end{document}